\documentclass[twocolumn]{aa}
\usepackage{graphicx}
\usepackage{psfig}
\usepackage{txfonts}
\newcommand{\vlos}{$v_{\mathrm{los}}$}
\newcommand{\los}{line-of-sight}
\newcommand{\vtang}{$v_{\mathrm{tang}}$}
\newcommand{\vrad}{$v_{\mathrm{rad}}$}
\newcommand{\vc}{$v_{\mathrm{circ}}$}
\newcommand{\kms}{$\mathrm{km\, s^{-1}}$}

\newcommand{\ak}{$A_{\mathrm K}$}

\newcommand{\Msun}{$\mathcal{M}_\odot$}

\newcommand{\lVd}{$lV$-diagram}

\newcommand{\rsun}{$r_\odot$}
\newcommand{\phisun}{$\phi_\odot$}

\newcommand{\rmin}{$r_{\mathrm{min}}$}

\newcommand{\wi}{$w_{\mathrm{i}}$}

\newcommand{\rgc}{$r_{\mathrm{GC}}$}

\newcommand{\vx}{$v_{\mathrm{x}}$}
\newcommand{\vy}{$v_{\mathrm{y}}$}
\newcommand{\xinit}{$x_{\mathrm{init}}$}
\newcommand{\yinit}{$y_{\mathrm{init}}$}
\newcommand{\vxinit}{$v_{x,\mathrm{init}}$}
\newcommand{\vo}{$v_{\mathrm o}$}
\newcommand{\vyinit}{$v_{y,\mathrm{init}}$}
\newcommand{\rinit}{$r_{\mathrm{init}}$}

\newcommand{\micron}{$\mu$m}
\newcommand{\ax}{$a_{\mathrm x}$}
\newcommand{\ay}{$a_{\mathrm y}$}

\newcommand{\yaxis}{$y-$axis}

\newcommand{\Phiaxisym}{$\Phi_{\mathrm{axisym}}$}
\newcommand{\Phibar}{$\Phi_{\mathrm{bar}}$}

\begin{document}
   \title{The distribution of maser stars in the inner Milky Way:\\
   the effect of a weak, rotating bar}
   \authorrunning{Habing et al.}
   \titlerunning{Maser stars and a rotating Galactic bar}
   \author{H.J. Habing, M.N. Sevenster, M. Messineo, G. van de
   Ven, K. Kuijken: the ROTBAR consortium \inst{1}}
   \offprints{H.J. Habing}
   \institute{Leiden Observatory,\\P.O. Box 9513,\\
    2300 RA Leiden,\\ the Netherlands
              \email{habing@strw.leidenuniv.nl}
   }
   \date{accepted by A\&A on july 26, 2006}
   
   \abstract{We derive the distribution of maser stars in the inner
     Milky Way based on an analysis of diagrams of longitude versus
     \los-velocity (= lV-diagrams) for two samples of maser stars: 771
     OH/IR stars and 363 SiO-maser stars. The stars are all close to
     the plane of the Milky Way and have longitudes from $-45^\circ$
     to $+45^\circ$.
     
     The two \lVd s are compared qualitatively and found to be very
     similar. They also compare well with the \lVd\ of interstellar
     CO, but there are significant differences in detail between the
     stellar \lVd s and that of the ISM.
     
     Based on the qualitative discussion we divide the \lVd s into
     seven areas. In each area we count the number of stars as
     observed and compare these numbers with those predicted by an
     assumed set of orbits in a galactic potential. This potential is
     axially symmetric but a weak rotating bar has been added. We
     conclude that the maser stars move on almost circular orbits
     outside of about 3.5 kpc, but that the orbits become more and
     more elongated when one goes deep inside our MW. We find a strong
     effect of the Corotation (=CR) resonance at 3.3 kpc, we see a
     small but noticeable effect of the Outer Lindblad Resonance at 5
     kpc and no effect of the Inner Lindblad Resonance (=ILR) at
     $r=0.8$ kpc.
   
     We find a set of 6 groups of orbits that together predict counts
     in agreement with the counts of stars observed. We then calculate
     the trajectory of each orbit and so find the distribution of the
     maser stars in the plane of the MWG.  This distribution has two
     new (but not unexpected) features. The first is a bar-like
     distribution within 2 kpc from the GC outlined. These orbits
     explain the high-velocity stars near $l=0^\circ$ in the forbidden
     and the permitted quadrants. The second feature are two
     "croissant''-like voids in the distribution close to the CR
     radius (3.3 kpc).  These voids are the consequence of the
     presence of the co-rotation resonance.
     
     We find excellent agreement with an earlier reconstruction by
     Sevenster (1999) based on partially the same data but on a
     completely different analysis.

   }

   \maketitle

%

\section{Introduction}

We cannot see into the inner parts of our Milky Way Galaxy (=MWG)
close to the galactic plane because of the extinction of light by dust
particles in intervening interstellar clouds. Radiation with a
wavelength of a few microns and longer is not affected by the dust and
that makes it possible to detect and to study the objects in the inner
Galaxy. In the 1950's, under the leadership of J.H.  Oort, the
distribution of the interstellar matter (=ISM) throughout the MWG was
derived in an elegant and convincing manner using the results of a
survey of the 21 cm line of HI. The success was not complete, however:
about one third of the galactic plane map was blank: in directions
within $30^\circ$ from the galactic center (=GC) and galactic
anticenter the kinematics of the gas could not be used to find the
distance to the gas velocities and in addition the gas motions in the
GC direction are complex and were not understood at that time. The
problem is apparently solved by the introduction of a weak, rotating
bar (see e.g.  Bissantz, Englmaier \& Gerhard, 2003).
\nocite{bissa:03} The 21 cm HI line data have been complemented in the
1970's by surveys of the lowest rotational transition of CO at 115 GHz
($\lambda = 2.6$ mm), a molecule that acts as substitute for molecular
hydrogen. A very conspicuous feature is the (almost complete) absence
of ISM inside the so-called ``molecular ring'' with a radius of about
3 kpc, except of the gas within a few hundred parsec of SgrA*.  All
these results are well known and have been described in text books-
e.g.  Binney \& Merrifield 1998 (hereafter BM) \nocite{binne:98}.\\

In 1968 Wilson and Barrett anounced the detection of a red giant as
the source of a strong emission line of the hydroxyl radical (OH) at
1612 MHz (18 cm).\nocite{wilso:68} The strength of the line by itself
and in comparison to other lines from the same multiplet indicated
that the line had been amplified by a maser process in the
circumstellar environment of this star. Quickly thereafter Elld{\'e}r,
R{\"o}nn{\"a}ng \& Winnberg (1969) \nocite{ellde:69} announced the
discovery of several similar maser sources in a systematic survey of
a small part of the Milky Way in Cygnus. These detections suggested
that a "blind" survey at 1612 MHz of a large fraction of the Milky Way
might find enough masers to derive the distribution of these stars in
the inner MWG, similar to what had been achieved for the ISM through
the 21 cm line. In the following years two more maser-amplified
molecular lines were detected: H$_2$O (Knowles et al., 1969
\nocite{knowl:69}) and SiO (Kaifu, Buhl \& Snyder, 1975)
\nocite{kaifu:75}.\\ 

The suggestion of a larger, blind 1612 MHz survey led to two searches
for maser stars along the northern Milky Way (Bowers, 1978a, 1978b;
Baud et al., 1981a). \nocite{bower:78a, bower:78b, baud:81a} Some 200
stars were detected and from the observed relation between their
longitude and latitude coordinate and their \los\ velocity\footnote{We
  will use the word \textit{line-of-sight velocity} instead of the
  more common term \textit{radial velocity} because the latter will be
  reserved for stellar motions to and from the GC} it appeared that
the stars form a thin, rotating disk with low density (a "hole")
within about 3 kpc from the GC, much like the hole in the ISM.  An
unexpected result was the discovery of a star with a high
\textit{negative} velocity ($-341$ \kms ) at a \textit{positive}
longitude $(l,b)= (0.3^\circ,-0.2^\circ$) (Baud et al., 1975).
\nocite{baud:75} More high-velocity stars were discovered later.  They
play an important role in this paper.\\ 

This paper deals with a renewed search for the galactic distribution
of maser stars, using sources detected in two recent surveys, one in
the OH 1612 MHz line, the other in the 86 GHz line of SiO. The data
obtained in these surveys have already been analyzed into quite some
detail- as we will discuss below.  Nevertheless we felt that further
analysis might still yield some interesting results- hence this
paper.\\ 

The paper is structured as follows. (1) We discuss the astrophysical
properties of the maser star population. (2) A summary is given of
what is already known about the relation between maser stars and the
bar of our MWG. (3) The two diagrams of longitude versus line-of-sight
velocity (=\lVd s) of the maser stars are compared qualitatively one
with another and then with that of interstellar CO.  (4) After
adopting a galactic potential we calculate possible orbits for the
maser stars in the galactic plane and their projection on the \lVd .
(5) Using the \lVd\ of the stars as a guide we select the orbits in
the galactic plane that the maser stars are likely to follow. (6)
Using these orbits we derive the spatial distribution of the maser
stars in the inner MWG.\\ 

\section{Physical properties of maser stars}

What stars show maser emission?  An answer may be found in the reviews
edited by Habing \& Olofsson (2003).  \nocite{habin:03} When stars
with a mass below 6 or 7 \Msun\ turn off the main sequence they will
quickly go through a number of different appearances. In the last
phase in which nuclear burning is still significant, the Asymptotic
Giant Branch phase (=AGB), the star is very luminous, very red and it
loses mass at a high rate but at a low velocity. The out-flowing
matter is seen as a cold circumstellar envelope and it is in such
envelopes that the OH, H$_2$O and SiO masers are produced. The
AGB-phase is short-lived; perhaps it lasts as short as 50,000 yr:
maser stars are rare, but they represent ordinary stars of low and
intermediate mass.  Model calculations predict that the luminosity of
an AGB star is tightly correlated with the age and the metallicity of
the star. It is still too early to exploit this correlation but one
important conclusion can already be drawn: maser stars represent
objects of an age (0.2 to 2 Gyr) that is intermediate between the age
of RR Lyrae and globular clusters (10 Gyr) and that of OB-stars and
open clusters ($\leq 0.2$ Gyr).\\ 

Almost all OH/IR stars in the Milky Way have been found within
$50^\circ$ from the GC in spite of deep searches farther away-
e.g. te Lintel Hekkert et al. (1991)\nocite{telin:91}: there are
very few maser stars known outside the solar circle, and those few
are probably of high mass. This rareness of maser stars in the
outer MWG can be explained by a lower metallicity: when stars of
low metallicity reach the AGB phase they develop into carbon stars
and there are not enough free O-atoms left in the circumstellar
envelope to form the OH radical or the SiO molecule in sufficient
numbers. If a maser is seen only when the stellar metallicity is
high enough, galactic metallicity introduces a bias in the samples
of detected maser stars.\\

Concerning terminology: The OH-maser stars discussed here have all
been detected in the 1612 MHz line, and will hence be called
"OH/IR stars". Stars detected because of their SiO maser emission
will be called "SiO-maser stars".

\section{Existing Surveys for maser stars; previously derived conclusions}

We have at our disposal a sample of 1134 maser stars (771 OH, 363 SiO
masers), all at low galactic latitudes ($\mid b\mid\leq 3.5^\circ$)
and in the inner MWG ($ |l|\leq 45^\circ$).The OH/IR sample contains
all OH/IR maser stars with a peak flux density above 0.5 Jy in the
area specified and with a \los\ velocity within $300$ \kms\ from zero
velocity in the Local Standard of Rest; see Sevenster et al. (1997a,
1997b, 2001) \nocite{seven:97a, seven:97b, seven:01}. The SiO maser
sample has been obtained by pointing the telescope at infrared objects
from the ISOGAL survey (see Messineo et al., 2002)\nocite{messi:02}.
Not all candidate infrared objects have been observed and the
longitude distribution may not be representative; all infrared objects
were within $1^\circ$ from the galactic plane. The OH- and the SiO-.
survey had a sensitivity limit low enough to find stars beyond the
GC.\\ 

We limited ourselves to these two surveys because we had all the data
readily available and because we understood the selection criteria and
thus the completeness of the surveys. We therefore did not include the
large sample of SiO-maser stars detected by Japanese astronomers, e.g.
Deguchi et al. (2004b) \nocite{deguc:04b} who used the IRAS catalogue
as the source of candidate stars, and this catalog is quite incomplete
near the galactic plane. For the same reason we did not use the
earlier OH/IR survey by te Lintel Hekkert et al. (1991).
\nocite{telin:91} One other limitation concerns the immediate
surroundings of the GC, roughly the area within $1.5^\circ$ from Sgr
A*. Results from deeper OH surveys in this area are available
(Lindqvist, Habing \& Winnberg, 1992a, Lindqvist et al. 1992b
\nocite{lindq:92a,lindq:92b} and Sjouwerman et al., 1998)
\nocite{sjouw:98} but as we believe that the structure in this area
has its own difficulties that are not of importance for the remainder
of the MWG we did not discuss these surveys and leave those for a
future paper.\\ 

Consider now the conclusions already reached concerning the
distribution in position and in velocity of the maser stars as far
as these are relevant for our present study.

\begin{enumerate}
\item Sevenster et al. (2000) \nocite{seven:00} calculated orbits of
stars in a three-dimensional MWG model under the assumption that
there are two integrals of motion, the energy, $E$, and the
angular momentum around the $z$-axis, $J$. A quite significant
conclusion follows from their figure 6: the distribution of the
angular momenta changes gradually at energies $E$ corresponding to
circular orbits with a radius $r= 2.0$ kpc. At higher energies
(i.e. farther outside in the disk) the stars move all on
almost-circular orbits and in the clock-wise direction as seen
from the galactic north pole. At lower energies and thus deeper
inside the MWG the distribution of the angular momenta becomes
flat: for a given energy all angular momenta have the same
probability and there are as many circular orbits in the
anti-clock-wise direction as in the clock-wise direction.
\item If in a galaxy with a rotating bar the density distribution
is described as a function of polar coordinates $(r,\phi)$ the
dependence on time must be such that $\phi=\phi_0-\Omega t$.
Tremaine and Weinberg (1984) \nocite{trema:84} have shown that as
a consequence of this symmetry the value of $\Omega$ can be
derived from the measurement of \los\ velocities in an edge-on
galaxy. Using a data base of OH/IR stars almost the same as ours
Debattista, Gerhardt \& Sevenster (2001)\nocite{debat:02}
calculated the pattern speed of a bar in the inner MWG  and
obtained the value $\Omega= -59\,\pm 5$ km s$^{-1}$ kpc$^{-1}$.
The tangential velocity of the maser stars thus contains a
systematic component that is proportional to the distance of the
star to the GC. This is a strong argument in favor of the
statement that the OH/IR stars are part of the galactic bar.
\item Sevenster (1999b) \nocite{seven:99b} analyzed a sample
of 509 OH/IR stars in the longitude range from $l=+10^\circ$ down
to $l=-45^\circ$. She concludes: "Unequivocal morphological
evidence is presented for the existence of a central Bar...".
Corotation (CR) is at 3.5 kpc, the inner Lindblad resonance (ILR)
is between 1 and 1.5 kpc and the inner ultra-harmonic radius is at
2.5 kpc.

\item Sevenster, Saha, Valls-Gabaud \& Fux (1999)
\nocite{seven:99a} have compared several $N$-body simulations of
the MWG with the OH/IR stars data obtained with the ATCA. In these
simulations a bar is formed spontaneously. Good agreement between
predictions and observations is found for a viewing angle of
$44^\circ$, CR at 4.5 kpc.

\item In the same paper as in item 3 Sevenster points out
that the distribution of the OH/IR stars over positive and
negative longitudes are slightly different and that this
difference has to be attributed to the presence of the bar. We
have now a larger data base and our conclusion confirms the one
previous; see the next subsection.
\end{enumerate}

We conclude that all these items, except 1, argue that the maser
stars are part of the bar of the MWG.

\subsection{Differences between the distribution at positive and
at negative longitudes}

   \begin{figure}\label{fig:lplusminlmin}
    \includegraphics[width=6cm, angle=270]{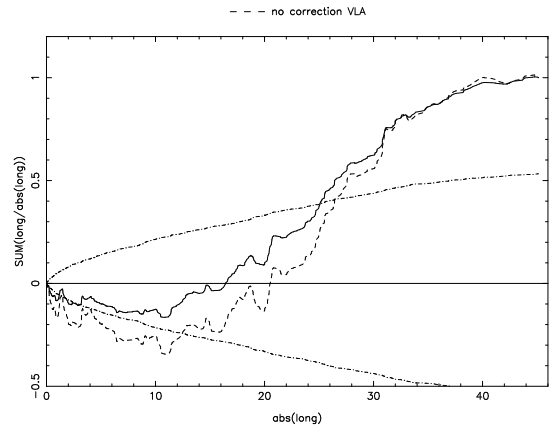}
    \caption{The cumulative distribution of the sign of the longitude,
     as a function of absolute longitude.The solid curve indicates
     the distribution when correcting for the lower completeness of
     the VLA section of the survey (see Sevenster et al 2001). The
     dashed curve indicates the distribution when no correction is
     applied, i.e. when effectively every star in the sample counts as
     1. A decreasing sum indicates an overdensity at negative
     longitudes, an increasing sum indicates an overdensity at
     positive longitudes.  Horizontal sections in the curve indicate a
     symmetrical distribution of stars with respect to l=0. The
     dash-dotted curves give the two-sided 95\% interval assuming the
     stars are drawn from a binomial distribution. Only OH/IR stars
     (614 in total) with outflow velocities between 2 and 20 km/s are
     included, to ensure a homogeneous population of stars that traces
     the Bar}
   \end{figure}

For a long time, the interpretation of asymmetries in the
integrated-light distribution observed toward the inner Galaxy was
limited by the degeneracy of the deprojection. In external
galaxies, edge-on bars do not show up as asymmetries in the
morphology, so asymmetries have to be interpreted as $m=1$
distortions (or odd-numbered in general). In the Galaxy, both
$m=1$ and $m=2$ non-axisymmetries would result in an asymmetric
surface morphology. Most observations before COBE (Weiland et al.
1994, Dwek et al. 1995)\nocite{weila:94,dwek:95} showed
asymmetries that could not be unambiguously attributed to an $m=2$
distortion.\\

The COBE observations themselves showed a surface-density distribution
with a clear hint of the Bar. The image is, however, still limited by
extinction. As explained by Blitz \& Spergel (1991)\nocite{blitz:91},
when observing a tracer population that can be sampled out to $d>12$
kpc, so that the bar is fully included, a signature would be seen that
would point unambiguously toward the existence of a bar. This
signature was first seen in a sample of OH/IR stars (Sevenster 1999).
With the extended sample of OH/IR stars, we now repeat the test.\\ 

Figure~1 shows this signature. After arranging all OH/IR stars
according to their absolute longitude we show $N=\Sigma l_i/|l_i|$. If
the distribution were axisymmetric $N$ would always remain close to 0.
This is clearly not the case.  Out to about $|l|=10^\circ$ there is an
overdensity at negative longitudes. Between $|l|=10^\circ$ and
$30^\circ$ there is a strong overdensity at positive longitudes. At
higher longitudes, the curve approximates a constant value, indicating
a certain lopsidedness in the distribution of stars in our MWG. For a
more elaborate discussion of this test, see Sevenster (1999)
\nocite{seven:99b}. An important conclusion for this paper is that the
sample of OH/IR stars does sample the entire Bar, both the near and
the far end. It provides a good sample for kinematical analysis of bar
orbits, especially with the accurate line-of-sight velocities.\\ 

The same asymmetric distribution of stars has been found by
Benjamin et al. (2005) \nocite{benja:05} in a much larger sample of red
giants detected at mid-infrared wavelengths with the SPITZER
satellite; the authors also argue for the existence of the bar.\\

\subsection{Foreground extinction and distance from the Sun}

Stellar masers are formed in the dust-rich envelopes of AGB stars.  In
OH masers these envelopes are optically thick to wavelengths as large
as 10 \micron . SiO masers form in thinner envelope and if near-IR
photometry is available one can derive \ak , the sum of interstellar
and circumstellar extinction of each maser star. As the maser stars
are variables one needs simultaneously measured magnitudes. Existing
surveys provide this information: $ I- ,J-$ and $K-$magnitudes by
DENIS and $J- , H-$ and $K-$magnitudes by 2MASS.  In all fields the
maser star is (one of) the brightest sources and can easily be
identified (Messineo et al., 2004, 2005)\nocite{messi:04,messi:05}.
When we separate groups of maser stars based on their value of \ak\ we
see systematic differences between their \lVd s; see Fig.~2. These
differences we attribute to differences in distance: larger values of
\ak\ imply a larger distance. The few stars at longitudes larger than
$20^\circ$ and with medium or large extinction may be located behind
dense clouds . The systematic change in character between the three
diagrams is in agreement with the model that we will develop later in
this paper.

   \begin{figure}\label{fig:AvSiOstars}
   \centering
   \includegraphics[width=6cm]{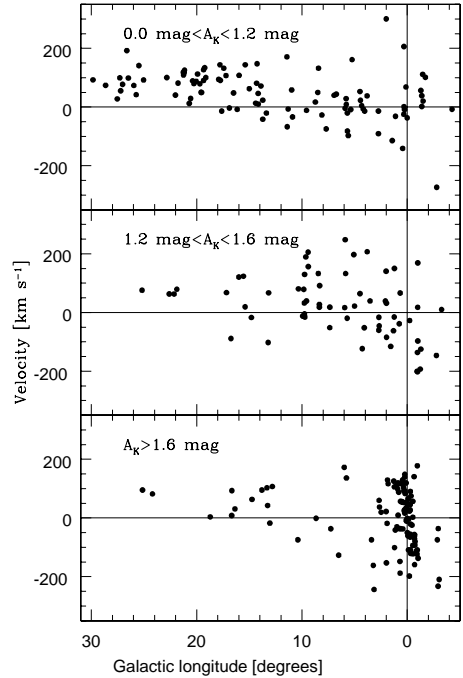}
   \caption{\lVd s of different samples of SiO-maser stars; the
   samples have been selected using an estimate of the amount of
   the foreground extinction}
   \end{figure}

\section{\lVd s of maser stars and of IS gas: a qualitative comparison}
\label{sec:qualitative}

We will compare the two \lVd s of the maser stars with each other and
with the \lVd\ of interstellar CO. The \lVd s are divided into four
quadrants in the usual way, see the upper diagrams in Fig.~3. Q2 and
Q4 are the well-known "forbidden quadrants" of circular galactic
rotation. Q1 and Q3 will be called the "permitted quadrants".\\ 

\subsection{Comparing the \lVd s of the maser stars}

The \lVd\ of the OH/IR stars is the upper left diagram in Fig.~3 and
that of the SiO-maser stars is the upper right diagram. The two
samples of maser stars have very similar distributions:
\begin{enumerate}
\item The \vlos$=0$ line is an approximate lower limit for \vlos\ at
positive longitudes and an upper limit at negative longitudes;
there are a few stars outside of this approximate limit.
\item The curve \vlos$=200\, (1 - \sin\, l)$ \kms\ is an upper limit in Q1
and \vlos$=-200\, (1 + \sin\, l)$ \kms\ a lower limit in Q3 for
most of the stars. The longitude distribution of the SiO maser
stars is not representative for the true distribution, but the
distribution in \vlos\ \textit{is} representative.
\item For $25^\circ \geq l \geq -25^\circ$ the curve \vlos$= 284\,
\sin\, l$ \kms\ separates a region of lower density of stars from
one of higher density.
\item Close to $l=0^\circ$ there is a significant number of stars with
$\mid$\vlos$\mid > 200$\,\kms .
\end{enumerate}

A clear difference between the two \lVd s is the very tight
correlation between $l$ and \vlos\ in the SiO-maser stars near
$l=0^\circ$; in the OH/IR diagram such a sample is not present-
although a trace of it may be seen by a favorably inclined reader.
The lack of a similar set of OH/IR stars is only apparent; in fact, in
an earlier, deeper VLA study of the OH/IR stars in a small region
around the GC (Lindqvist, Winnberg, Habing and Matthews, 1992a,
Lindqvist, Habing and Winnberg, 1992b)\nocite{lindq:92a,lindq:92b} a
sample of OH/IR stars was found with the same narrow correlation
between $l$ and $V$. The correlation may be explained by solid body
rotation caused by a \textit{constant} density in the mass
distribution inside 150 pc from the GC (Lindqvist et al.  (1992b)
\nocite{lindq:92b}, but another interpretation of the same data is
certainly possible (Winnberg, 2003).  \nocite{winnb:03}

\subsection{The \lVd\ of interstellar CO}

In the lower left corner of Fig.~3 we show the CO \lVd\ by Dame et al.
(2001).\nocite{dame:01}

\begin{enumerate}
  
\item The CO emission is strongly confined to the galactic plane, much
  more so than the maser stars.

\item The CO emission is (almost completely) limited to the permitted
  quadrants Q1 and Q3 and by the two curves defined by \vlos$< 200\,
  (1- \sin l)\,$\kms\ for $l > 0^\circ$ and by \vlos$> -200\, (1+\sin
  l)\,$\kms\ for $l < 0^\circ$. When the gas follows circular orbits
  these curves indicate the maximum \vlos\ in the direction $l$.

\item There is some emission in the forbidden quadrants Q2 and Q4.
  (i) Close to $(l,V)=(0,0)$ a weak emission reaches measurable
  intensity in Q2; its border is given by a straight line through
  $(l,V)=(0,-57)$ and $(l,V)=(15,0)$. In Q4 the same is true; this
  time the border line runs through $(l,V)=(0,25)$ and
  $(l,V)=(-15,0)$. (ii) There is weak emission in the forbidden
  quadrants farther away from the center; this emission is almost
  certainly from gas outside of the solar circle at the far end of the
  MWG.
  
\item A curve defined by \vlos$=280\, \sin l\,$\kms , separates a
  region of high intensity and one of low intensity as well as in Q1
  as in Q3; this is true both at positive and at negative longitudes
  in the \lVd\ of the OH/IR stars and of the SiO-maser stars (Fig. 3).
  This curve is the outer limit of the large and well-known hole in
  the ISM distribution inside of the molecular ring. If the ISM moves
  in perfect circles around the GC at a constant speed \vc\ the factor
  $280.4\,$\kms\ is fixed by the radius of the hole: $280.4=
  v_{\mathrm o} ( r_{\mathrm{GC}} /r_{\mathrm{hole}}-1)$ and because
  we use (see below) \rgc $=8.0$ kpc and \vo $= 200$\,\kms we find
  $r_{\mathrm{hole}} = 3.3$ kpc.
  
\item Two straight lines defined by $ l =\pm 6^\circ$ limit the region
  of very high velocities around $l=0^\circ$. The most extreme
  velocity is 250 \kms\ in Q1 and -250 \kms in Q3. High velocities are
  not found in the forbidden quadrants Q2 and Q4.
\end{enumerate}

\subsection{Comparing the stellar and the ISM \lVd s}

The comparison of the two maser star \lVd s with each other shows
these to be very similar. This is almost, but not quite the case
when one compares the maser star \lVd s with that of interstellar
CO.\\

\begin{enumerate}
\item The velocities of the stars are limited by the same curves and
  straight lines as the velocities of the gas but a small number of
  stars have wandered outside these curves.
\item There is quite some "trespassing" by the maser stars into the
  two forbidden quadrants Q2 and Q4, much more than shown by the ISM.
\item Our sample of OH/IR stars contains 28 stars in the strip between
  $l=6^\circ$ and $l=-6^\circ$ with a velocity $\mid$\vlos$\mid \geq
  200$\,\kms\ and four SiO-maser stars. Nineteen OH/IR stars have
  "permitted" velocities (i.e. they are found in Q1 and Q3) and nine
  have "forbidden" velocities (i.e. they are found in Q2 and Q4);
  there is no high-velocity gas at forbidden velocities.
\item The curve \vlos$=280\, \sin l\,$\kms separates a region of high
  source density and one of low source density in both Q1 and Q3.
\end{enumerate}

\subsection{Conclusions based on our qualitative comparison}

The stellar \lVd s show a small but significant number of stars
outside the limits strictly observed by the CO. We attribute this
to the slow but continuous diffusion of stars in velocity space:
stars accumulate perturbations of their velocity by incidental
encounters, whereas the kinematics of the gas is dominated by
dissipative forces. \\

\begin{figure*}\label{fig:vierdiagrammen}
   \centering
   \centerline{
   \psfig{figure=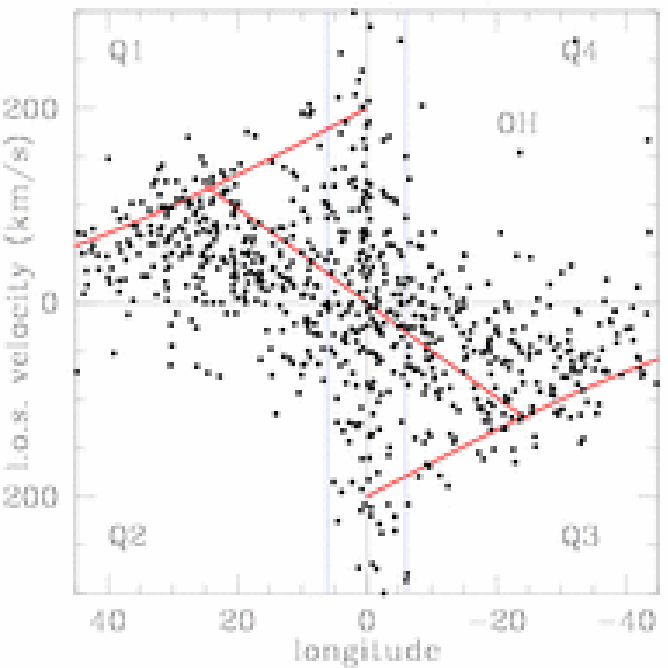,width=6cm}
   \psfig{figure=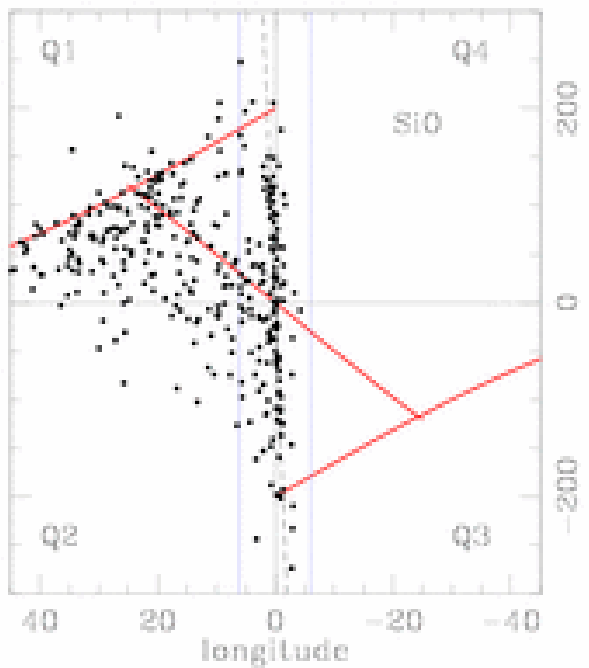,width=6cm}}
   \centerline{
   \psfig{figure=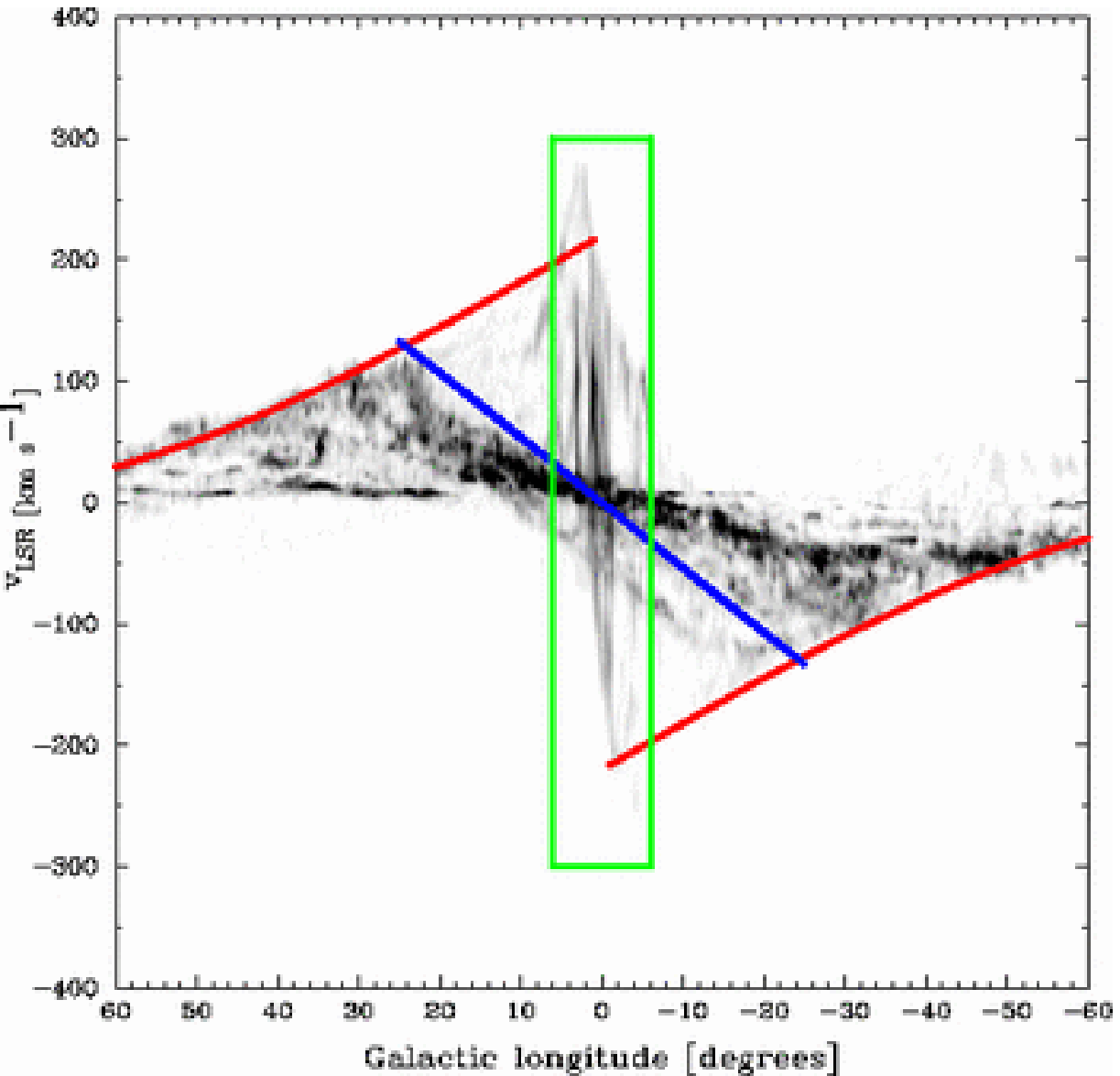,width=6cm}
   \psfig{figure=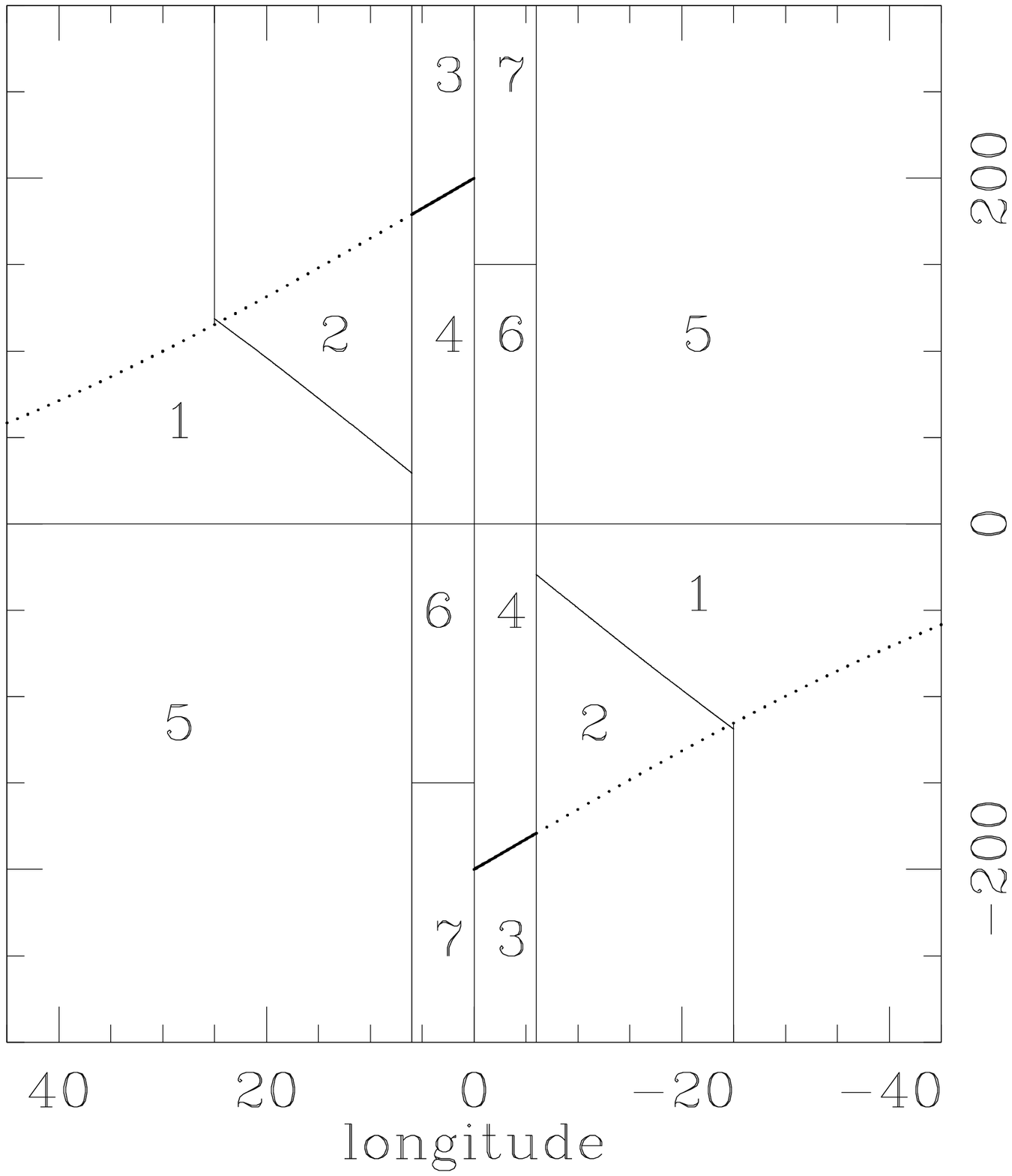,width=6cm}}
   \caption{The two diagrams at the top show the distribution of
     longitude and l.o.s. velocity of OH/IR stars (left) and SiO maser
     stars (right). The meaning of the red curves in these two
     diagrams is explained in the text. The dashed line in the SiO
     diagram is the line of correlation between longitudes and
     velocities of stars very close to the galactic center: $V=150\,
     l\,$\kms\ where $l$ is in degrees. The diagram of the lower left
     shows the diagram for the interstellar CO emission (Dame et al.,
     2001). The diagram at the lower right shows the
     limits of the areas defined in the text. The dotted curves are
     porous in the sense that they are a limit to most stars but not 
     to all}
\end{figure*}

All in all we conclude that \lVd s of the maser stars agree with
those of the ISM in their major features. The agreement thus
implies that stars and gas have a very similar distribution in
space and in velocity.\\

Our next step is to analyze the two stellar \lVd s by a quantitative,
statistical method. Based on our qualitative comparison we divide the
\lVd s into seven areas- see the diagram on the lower right in Fig.~3.
We count the OH/IR stars in each area and compare the count to what we
predict with a simple galactic potential to which a weak rotating bar
is added. Each area has the same weight although different areas
contain different numbers of stars. For example the small number of
stars in area 7 carry the same weight as the much larger number of
stars in area 1.

\section{A quantitative analysis of the \lVd s of the stars}

Our quantitative analysis of the stellar \lVd s is based on the orbits
of stars calculated in a simple two-dimensional gravitational
potential that contains a weak, rotating bar. From this potential we
generate a library of orbits representative for our problem.  Each
orbit defines a probability distribution in the \lVd s. By taking
linear combinations of orbits we predict numbers of stars in several
areas in the \lVd s. We then use a $\chi^2$-method to find the best
agreement between predictions and observations.\\ 

\subsection{Calculation of orbits; equations of motion}

We start from the equations given in the monograph by Binney \&
Tremaine (1987, hereafter BT)\nocite{binne:87}. We adopt a coordinate
system corotating with the bar and with its origin at the center of
the MWG. We start from the equation
\begin{equation}
  \label{eq:eqofmotion}
  \ddot{\mathbf{r}} = -\nabla\Phi - 2\left(\mathbf{\Omega_b}
  \times\dot{\mathbf{r}}\right).
\end{equation}
and obtain as equations of motion:
\begin{equation}
    \label{eq:d2xdt2}
{{d^2 x}\over{dt^2}}= -{\partial\Phi\over\partial x}
+\Omega^2\,x+2\, \Omega\, {{dy}\over{dt}}
\end{equation}
 and
\begin{equation}
   \label{eq:d2ydt2}
{{d^2 y}\over{dt^2}}= -{\partial\Phi\over\partial y}
+\Omega^2\,y-2\,\Omega\, {{dx}\over{dt}}
\end{equation}
In the right-hand side of equations (\ref{eq:d2xdt2}) and
(\ref{eq:d2ydt2}) the first term is the gravitational acceleration
in the direction of the $x$-, resp. \yaxis; the second term is the
centrifugal acceleration and the third term is the Coriolis
acceleration. The Coriolis acceleration introduces a dependence of
the acceleration in the $x$-direction on the velocity in the
$y$-direction and vice versa.\\

We solved equations (\ref{eq:d2xdt2}) and (\ref{eq:d2ydt2})
numerically with the fourth order Runge-Kutta algorithm described in
"Numerical Recipes". We tested the code successfully by calculating
the same circular orbit in frames rotating at different speeds
including speed zero. The calculation of the Jacobian energy (see BT)
was another test; this energy was constant to within 1 in 1000.\\ 

The next step is to calculate the longitude $l$ of the star and its
\los\ velocity \vlos\ at a given moment in time $t$. Let the cartesian
coordinates of the star be ($x,y$) and its polar coordiates
($r,\phi$); the corresponding coordinates of the Sun are
($x_\odot,y_\odot$) and (\rsun,\phisun). The coordinate \phisun\ is
the angle between the long axis of the bar and the Galactic Center -
Sun direction. The parameter \phisun\ is one of the important free
parameters of our problem. Consider the triangle of the Sun (symbol
Z), the star (symbol S) and the galactic center (symbol C).  The angle
SCZ will be called $\beta$; its value is $\beta=$ $\phi-\phi_\odot$.
The distance $d$ from the Sun to the star is given by $d^2=r^2+
r_\odot^2-2\ r\ r_\odot \cos\ \beta$.  The longitude $l$ is derived
from the equation $\sin(-l)/r=\sin \beta /d $; the minus sign in front
of $l$ takes into account the definition of longitude in the
counterclock direction when viewed from the galactic north pole. For
the calculation of \vlos\ we have to add the velocity of rotation of
the frame and thus obtain the velocity of the star in an inertial
frame that coincides with the rotating frame at the time $t$: $(\dot
x_{\mathrm{in}},\dot y_{\mathrm{in}})= (\dot x- r\, \Omega \,\sin\,
\phi, \dot y+ r\,\Omega\, \cos\, \phi)$.  The projection of the
velocity of a given star on the \los\ equals $v_{\mathrm{los},1}= \dot
x_{\mathrm{in}} {({x-x_\odot})/ d}+ \dot y_{\mathrm{in}} {({y-y_\odot
  })/ d}$. The projection of solar velocity on the \los\ equals
$v_{\mathrm{los},2} =v_\odot \sin\, l$ and \vlos\ is the sum of the
two projections: $v_{\mathrm{los}}= v_{\mathrm{los},1}+
v_{\mathrm{los},2}$.\\ 

The observations tell us the number of stars in well defined areas in
the \lVd s. In our calculations we follow a star in an orbit defined
by initial position $\vec r_0$ and an initial velocity $\vec{\dot
  r_0}$ and record the time that the star spends in each such area. We
followed the orbit over a time long enough for the star to make
several turns around the galactic centre. We checked that if we
increased this time by a factor 2.5 (50,000 time steps instead of
20,000) the relative time spent in each area did not change
significantly.\\

At a certain time $t$ all stars have the position
($r_{\mathrm{rot}},\phi_{\mathrm{rot}}$) in the rotating frame and the
position ($r_{\mathrm{in}},\phi_{\mathrm{in}}$) in an inertial frame
that coincided with the rotating frame at $t=0$.  Because
$r_{\mathrm{in}}=r_{\mathrm{rot}}$ and $\phi_{\mathrm{in}}=
\phi_{\mathrm{rot}}+\Omega t$ the distribution of the stars in the
rotating frame at time $t$ is equal to that in the inertial frame
after rotating the former distribution over an angle $\Omega t$.

\subsection{The gravitational potential}

We adopt a simple potential taken from BT (their eq. 3.77):
\begin{equation}
  \label{eq:Phipotbar}
  \Phi_{\mathrm{bar}}(x,y) = \frac12 v_0^2 \ln \{ r_0^2+x^2+(1+\epsilon)y^2\}.
\end{equation}
The coordinate frame that underlies this potential rotates with an
angular speed of -60 \kms kpc$^{-1}$. When $\epsilon=0.0$ the
potential is axially symmetric:

\begin{eqnarray}
  \label{eq:Phipotaxisym}
\Phi_{\mathrm{axisym}}(r,\phi)={1\over 2}\, v_{\mathrm{o}}^2\
\ln\,\{r_{\mathrm{o}}^2+ r^2\}
\end{eqnarray}
where $r=\sqrt{x^2+y^2}$. A possible rotation of the frame must
have no influence on the orbits of this axially symmetric potential. The
calculations bear this out; in fact, we used this independence on $\Omega$
as one of the tests on our software.\\

For later use we define a circular velocity, \vc , by using \vc$^2 =
(1/r)d\Phi/dr $ which leads to $v_{\mathrm{circ}}=v_{\mathrm o}\,{r/ 
  \sqrt{r_{\mathrm o}^2+r^2}}$.\\

Eq.~(\ref{eq:Phipotaxisym}) is mathematically easy to use and it
satisfies two important boundary conditions: (1) close to the galactic
center the circular velocity increases linearly with $r$, as had been
deduced from the narrow correlation between $l$ and \vlos :
$v_{\mathrm c} = v_{\mathrm o} (r/r_{\mathrm o})$\ \kms (see above);
  (2) the circular velocity is effectively constant beyond 0.5 kpc.
  The correlation between $l$ and \vlos\ of SiO maser stars disappears
  for $\mid l \mid \geq 1^\circ$; therefore we
  will use $r_o = 150$ pc ($=r_\odot\sin 1^\circ$) and \vo$=200$ \kms .\\

Eq.~(\ref{eq:Phipotbar}) is also easy to handle. We chose
$\epsilon=0.1$, a value of the same magnitude as that of the barred
potential used by Bissantz, Englmaier \& Gerhard (2003)
\nocite{bissa:03} for their interpretation of the structure of the ISM
in the inner MWG. This low value of $\epsilon$ guarantees that the bar
has only a weak effect: in any point $(x,y)$ the gravitational
accelerations in the $x$- and $y$-directions, \ax\ and \ay\ 
respectively, have a ratio $a_x/a_y \geq 0.89\, x/y $, i.e. only 11 \%
lower than in the axially symmetric case. As a consequence, the
rotating bar has a limited effect on the actual stellar orbits in
general but may have much larger effects near a resonance. The value
of $\epsilon$ is quite uncertain- see for some discussion Sect. 6.3.\\

\subsection{Orbital structure}

\subsubsection{Non-conservation of angular momentum}

The properties of the stellar orbits are best discussed in the polar
coordinates, ($r,\phi$), rather than in cartesian coordinates $(x,y)$.
In polar coordinates the equations take the following form

\begin{eqnarray}
  \label{eq:Phipolar}
\Phi(r,\phi)={1\over 2}\, v_{\mathrm{o}}^2\
\ln\,\{r_{\mathrm{o}}^2+ r^2\,(1+ \epsilon\,\sin ^2\phi)\}
\end{eqnarray}
and
\begin{eqnarray}
  \label{eq:firstrphi}
{{d^2 r}\over {dt^2}}= -{{v_o^2}\over {r}}{{1+\epsilon \sin^2
  \phi}\over{r_o^2/r^2+1+\epsilon\sin^2\phi}} + {{J^2}\over
  r^3}
\end{eqnarray}
and
\begin{eqnarray}
      \label{eq:secondrphi}
  {{dJ}\over{dt}}= - {1\over 2} v_o^2 {{\epsilon \sin 2
  \phi}\over {r_o^2/r^2+ 1+\epsilon \sin^2 \phi}}
\end{eqnarray}
where $J=r^2 (\dot\phi + \Omega)$, i.e. $J$ is the angular
momentum in the inertial frame.\\

Both potentials, $\Phi_{\mathrm{bar}}$ and $\Phi_{\mathrm{axisym}}$
are monotonously rising when $r$ increases and thus all orbits are
bound. The first term in the right-hand side of
Eq.~(\ref{eq:firstrphi}) equals the gravitational attraction ($=d\Phi
/ dr$) and the second term equals the centrifugal acceleration
($J^2/r^3$). If $J$ has a lower limit different from 0 there will be a
minimal radius, \rmin, where the centrifugal acceleration equals the
gravitational acceleration and where the radial velocity, \vrad , equals
zero.\\ 

Some more insight is found by rewriting Eq. (\ref{eq:secondrphi})
as follows:
\begin{eqnarray}
\dot\phi\,{{dJ}\over{d\phi}}= -{1\over 2}v_o^2 {{\epsilon
\sin\,2\phi}\over{r_o^2/r^2+1+{1\over 2}\epsilon(1-\cos 2\phi)}}
\end{eqnarray}
The right hand side is independent of $r$ when $r\gg r_o$; say, when
$r> 0.5$ kpc. Divide the $xy$-plane in four quadrants in the usual
fashion. Integrating this equation over any of the quadrants, the
integral will have the same absolute value as the integral over each
of the other three; only the sign is different: over Q1 and Q3 the
integral will be positive, over quadrants Q2 and Q4 the integral is
negative. A star gains as much angular momentum in a given quadrant as
it looses in the next; periodic orbits are possible. Only when large
non-linear variations occur in $\dot\phi$ the star will gain or loose
net angular momentum over one revolution.\\

\subsubsection{Co-rotation and Outer and Inner Lindblad Resonance}

The assumption of small variations in $J$ does no longer hold for
those orbits that meet the bar each time under the same
conditions: then the accumulation of effects due the weak force
may be large.\\

The first case is that of stars on a circular orbit with a velocity
equal to that of the bar, i.e. when \vc$=r \Omega$. This radius is
called the radius of co-rotation, $r_{\mathrm{CR}}$; in our
calculation $r_{\mathrm{CR}}$=200/\,60 kpc = 3.33 kpc. Such stars have
zero velocity in the rotating frame, the gravitational attraction
equals the centrifugal force and at first sight it looks as if the
star will remain in the middle of the bar when that rotates around the
GC. On closer look this circular orbit is unstable. Suppose that the
star moves a little bit inward, into the direction of the GC, without
changing its tangential velocity.  Then the star will slowly move
ahead of the bar; gravitational attraction by the bar will decrease
the angular momentum and the gravitational attraction will pull the
star farther inward. The opposite happens when the star receives a
small velocity in the outward direction: now the bar will accelerate
the star, the centrifugal force increases and pushes the star further
outside. Thus orbits starting just outside of co-rotation and orbits
starting just inside of co-rotation have the remarkable property
that, while they originate from two neighboring points at almost the
same velocity they separate over some distance when $x=0$. The red and
blue orbits in Fig.~4 illustrate this.\\ 

The area around the $y$-axis between the two orbits can be filled by
high angular-momentum orbits. See the green and black orbits in Fig.~4
that start at the $y$-axis at $y=\pm 3.3$ kpc with almost circular
velocity in the tangential direction and zero velocity in the radial
direction. These orbits are almost stationary in the rotating frame
and thus also in the \lVd ; the two locations in the $xy$-plane are
called ``Lagrangian points'' (BT). If the population of these orbits
is significant one should see a cluster of points in the \lVd : one at
$l=-25^\circ$ and \vlos$=-116$ \kms\ and the other at $l=+17^\circ$
and \vlos$=+78$ \kms . Such a cluster of sources is {\textit{not}} seen in
the two observed stellar \lVd\ diagrams and we must assume that the
orbits around the Lagrangian points are populated by at most a few
stars if at all. As a consequence one expect voids, areas without
stars, between the two orbits; the voids have the shape of the
crescent of
the Moon- again: see Fig.~4.\\

Another consequence is that the two points at $(x,y)= (\pm
r_{\mathrm{CR}},0)$ are points of contact between two regions that are
otherwise separated Through these points a flow of ISM can be
established, inward or outward.  Later we will come back to this
point.\\ 

   \begin{figure*}\label{fig:corotatie}
   \centering
   \centerline{
   \psfig{figure=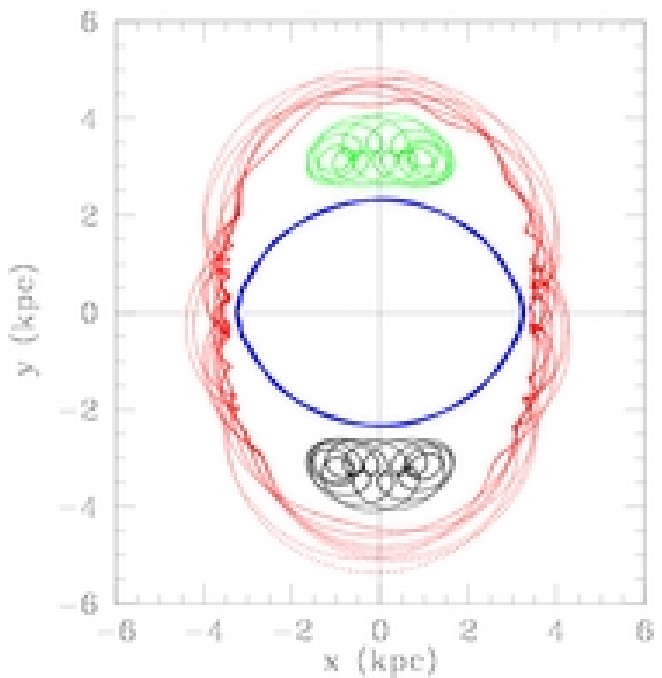,width=6cm}
   \psfig{figure=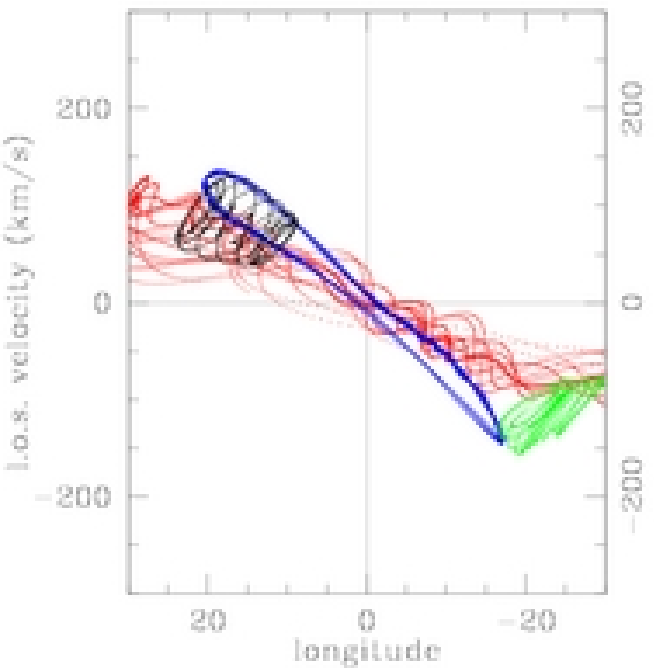,width=6cm}}
   \caption{Illustration of the co-rotation effect on almost circular
     orbits. The two orbits in red and in blue start at the center of
     the bar just inside and just outside co-rotation (at 3.25 and at
     3.40 kpc from the center) and with practically the same velocity
     in the $y$-direction (-199.8 \kms ). When they cross the $y$-axis
     they are separated over a distance of about 1.2 kpc. The green
     and black orbits start on the $y$-axis at 3.33 kpc and -3.33 kpc
     from the GC with a velocity of +179.8 and -179.8 \kms. The stars
     on either the green or the black orbit are practically stationary
     on the $xy$-plane and in the \lVd : see both diagrams. Stars on
     green and black orbits will cause a well-marked cluster of points
     in the stellar \lVd . Such clusters are not seen and thus the
     green and black orbits are not populated.}
   \end{figure*}

The second resonance is found for high-angular momentum orbits at
a radius of 5.0 kpc and this is 3/2 times the co-rotation radius.
This shows that we encounter there the Outer Lindblad Resonance
(=OLR). Its effects are mild. Orbits close to the OLR have a
larger extent in the $y$-direction than in the $x-$direction.\\

A third resonance, the Inner Lindblad Resonance (=ILR), is found
at $r=0.9$ kpc. It is caused by a resonance between the epicyclic
frequency and the period of rotation around the GC. The ILR is
irrelevant for our models, because in that region of the Milky Way
we use only orbits of small angular momentum and these are not
affected by the ILR.\\

\subsubsection{Low angular momentum orbits within 1 kpc from the
GC: rosette and bar orbits}

Low angular-momentum orbits within about 1 kpc from the GC turn out to
have an interesting property. The orbits have the shape made by a
succession of thin loops. Each loop is roughly symmetric with respect
to the GC; its very elongated shape (an axial ratio of 10 to 1 is
typical) is the result of an interplay between the gravitational
attraction and the Coriolis force; the centrifugal force turns out to
be unimportant. As expected, the bar has a minor effect on the shape
and size of the loop: compare the orbits (a) and (b) in Fig.~5 that
start at the same point, but while orbit (a) is calculated for the
axially symmetric potential ($\epsilon=0$), orbit (b) is calculated for the
barred potential ($\epsilon=0.11$). Orbits (b) and (c) have been
calculated for the same potential and they start at the same distance
from the GC but (b) starts on the $y$-axis and (c) on the $x$-axis.\\ 

  \begin{figure*}\label{fig:balkbaan}
   \centering
   \centerline{
   \psfig{figure=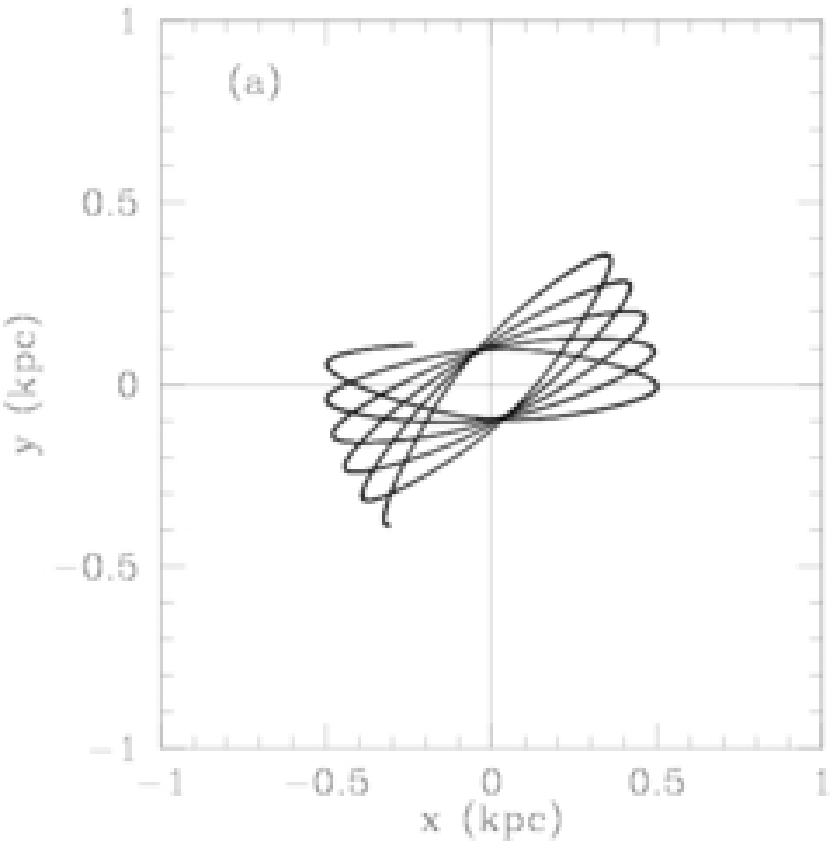,width=4cm}
   \psfig{figure=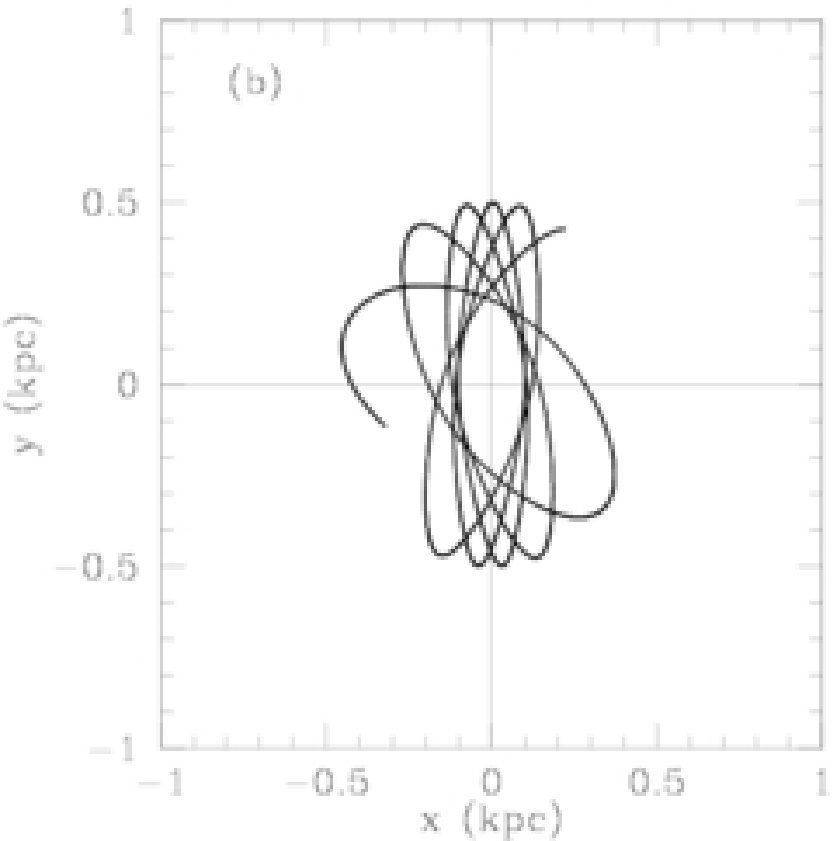,width=4cm}
   \psfig{figure=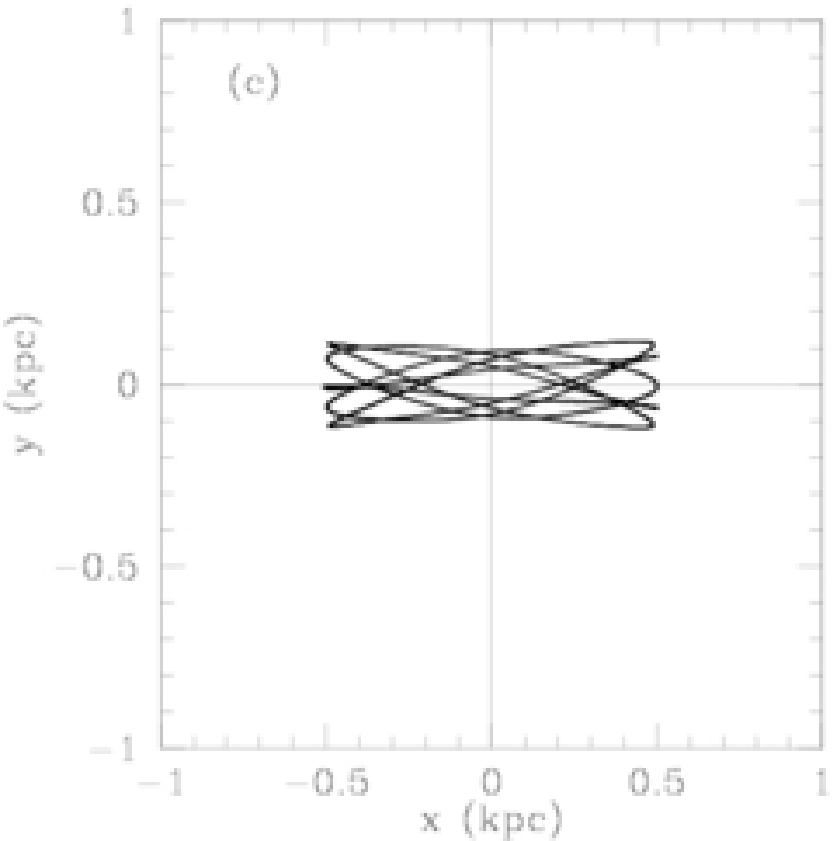,width=4cm}}
   \caption{Three comparable orbits. Orbit (a) has been calculated
     in the axially symmetric potential ($\epsilon = 0.0$) and started
     at a distance of 0.5 kpc from the GC with an initial radial
     velocity \vrad$=0$ and a tangential velocity \vtang$= 57.5$ \kms.
     Orbits (b) and (c) have been calculated for the bar potential
     using $\epsilon=0.11$.  Orbit (b) started at the $y$-axis with
     \vrad$=0$ and \vtang$=+57.5$ \kms\ and orbit (c) started at the
     $x$-axis at $x=0.5$ kpc and an initial velocity \vrad$=0$ and
     \vtang$=-57.5$\kms. For clarity we show only a small part of the
     orbit. Orbits of type c are called ``rosette-orbits'', orbits of
     type d are called ``bar-orbits''}
   \end{figure*}
   
   The bar-potential thus produces an interesting secular effect.
   Whereas orbit (b) describes a rosette around the GC, orbit (c) is
   restrained in the $y$-direction; its azimuth does not rotate
   around the GC. The star describes a rectangle:
   \textit{Orbits starting inside the bar remain inside the bar}.
   Orbits similar to orbit (c) will be called "bar orbits", orbits
   like orbit (b) will be called ``rosette orbits''.  We attribute the
   difference between bar and rosette orbits to the gravitational
   attraction by the bar.\\ 
   
   Bar and rosette orbits project differently in the \lVd - see the
   four diagrams at the top of Fig.~6. The rosette orbits cut the
   line-of-sight at a large variety of angles and \vlos\ has a large
   range of values. Stars on bar orbits travel up and down the bar and
   the line-of-sight always cuts the orbits at very similar angles;
   consequentially one finds high values for \vlos\ more often and
   they distribute more evenly among permitted and forbidden
   quadrants. This is illustrated in Table~1 and discussed in Sect.
   5.5. Orbits \#3 and \# 8 are the same as orbits in Fig.~5b and in
   Fig.~5c. Whereas orbit 8 spends essentially zero time in area 7,
   orbit 3 spends 10\% of its time there.\\

\subsection{Orbit library}

The orbits in \Phibar\ are characterized by the initial values for
$x$, $y$, \vx\ and \vy . To obtain a complete library of orbits one
needs to explore a four-dimensional space. This is a huge task and
there is the risk of using ``twin orbits'': two orbits that set out on
different initial conditions but one of the two passes through the
starting point of the other with the same velocity, so that the two
orbits are the same. To obtain an orbit library representative for our
sample we used a selection method that works very well for orbits
belonging to \Phiaxisym : label orbits by their initial distance,
\rinit , and adopt \vrad$=0$ and \vtang$=f\,$\vc, where $f$ is a
constant with values between 0 and 1.  By chosing \vrad$=0$ one
ensures that \rinit\ is the apocenter of the orbit.\\ 

To select orbits in \Phibar\ we chose various values for \rinit\ and
for $f$. For each pair (\rinit$,f)$ we started one orbit at $(x,y)=
($\rinit$,0)$ with velocity $(0,-f$\vc$)$ and a second at $(x,y)=
(0,$\rinit$)$ with a velocity $(f$\vc$,0)$; we used values for
\rinit\ between 0.50 to 7.50 kpc, and values for $f=0.3, 0.5, 0.85,
1.0$.\\

\subsection{The projection of orbits on the areas in the \lVd s}

The two upper \lVd s of Fig.~3 contain all our information on the
orbits of the maser stars. The importance of each dot in the diagrams
depends on where it is found and precisely this is the idea behind our
definition of the boundaries of the seven areas (as given in the
diagram on the lower right in Fig.~3): \textit{a priori} we judge each
area to be of the same importance. For example, area 7 contains only
12 stars and yet these stars are essential because the orbits on which
they move are not implied by the dots in any of the other areas. In
contrast, area 1 contains 355 stars that together represent all orbits
of high angular momentum outside of the molecular ring, i.e. the
population of the galactic disk.\\ 

Table~1 illustrates this quantitatively. Each row refers to an orbit
that is characteristic for its group. The full specification of each
orbit is given in Table~2, see farther below. Table~1 gives the
percentage of time of each orbit spends in each area (columns 3 to 9).
Column 10 contains the fraction of time spent outside the longitude
range considered, i.e. the time spent at $\mid l \mid\geq 45^\circ$.
It is easily seen that only stars on orbits of group 1 and 3 populate
area 7.\\

\begin{tabular}[b]{rr|rrrrrrrr}\hline
\multicolumn{10}{c}{Table~1}\\
\hline
&&\multicolumn{8}{c}{\% of time spent in each selected area}\\
gr.&orb.& 1& 2& 3& 4& 5& 6& 7& out\\
\hline
1&  3&  0 &  0 &  19 & 40 &  0 & 31 & 10 &  0\\
2&  8&  0 &  0 &   8 & 71 &  0 & 21 &  0 &  0\\
3& 16&  3 & 23 &  15 & 43 &  1 & 14 &  2 &  0\\
4& 23& 14 & 69 &   0 &  8 &  1 &  8 &  0 &  0\\
5& 29& 55 & 16 &   0 &  8 & 12 &  6 &  0 &  2\\
6& 38& 51 &  2 &   0 &  3 & 22 &  4 &  0 & 18\\
\hline
\end{tabular}\label{tab:orbitsandareas}

\subsection{The comparison of observations and models;
$\chi^2$-values}

We selected orbits from the orbit library and labelled each with
the suffix $i$. Each orbit defines a probability distribution in
the \lVd , $p_i(l,V)$. We calculated the integral of over each
critical area $j$ and thus predicted the number of stars,
$N_{\mathrm{pred,j}}$, in each area, $j$:
\begin{eqnarray}
    N_{\mathrm{pred,j}}= N_{\mathrm{total}}\,
    \Sigma_i \, w_i \,\int\int_j p_i(l,V)\,dl\,dV
\end{eqnarray}
where we introduced weights $w_i$ that will be assigned to each
orbit. The weight is thus proportional to the number of stars on
this orbit. Later we determined a multiplication factor W= 771 /
$\Sigma_i w_i$; 771 is the total number of OH/IR stars of our
sample. After starting the calculations it quickly turned out that
44 individual orbits is a too large number too handle conveniently
and we formed six groups of orbits and give the same weight \wi\
and the same value of $f$ to all members of a group.\\

We compared the numbers observed in each area, $N_{\mathrm{obs}}$,
with the numbers predicted, $N_{\mathrm{pred}}$. We assume that
the probability distribution of $N_{\mathrm{obs}}$ is Poisson-like
and thus that the dispersion in the observed numbers equals the
mean value. We therefore use the two statistics

\begin{eqnarray}
  \label{eq:defchi1}
\chi^2_1 = {{({N_{\mathrm{obs}}^{\mathrm{nm}}
-N_{\mathrm{pred}}^{\mathrm{nm}})^2}
\over{N_{\mathrm{pred}}^{\mathrm{nm}}}}}
\end{eqnarray}

and

\begin{eqnarray}
  \label{eq:defchi2}
\chi^2_2 = {{({N_{\mathrm{obs}}^{\mathrm{nm}}
-N_{\mathrm{pred}}^{\mathrm{nm}})^2}
\over{N_{\mathrm{obs}}^{\mathrm{nm}}}}}
\end{eqnarray}

Ideally one expects the values of $\chi^2$ to be 1. In our case,
when comparing observations with predictions, we adjust the values
of $w_i$ to minimize $\langle\chi^2\rangle$.\\

\begin{figure*}[t!]\label{fig:twaalfdiagrammen}
   \centering
   \centerline{\psfig{figure=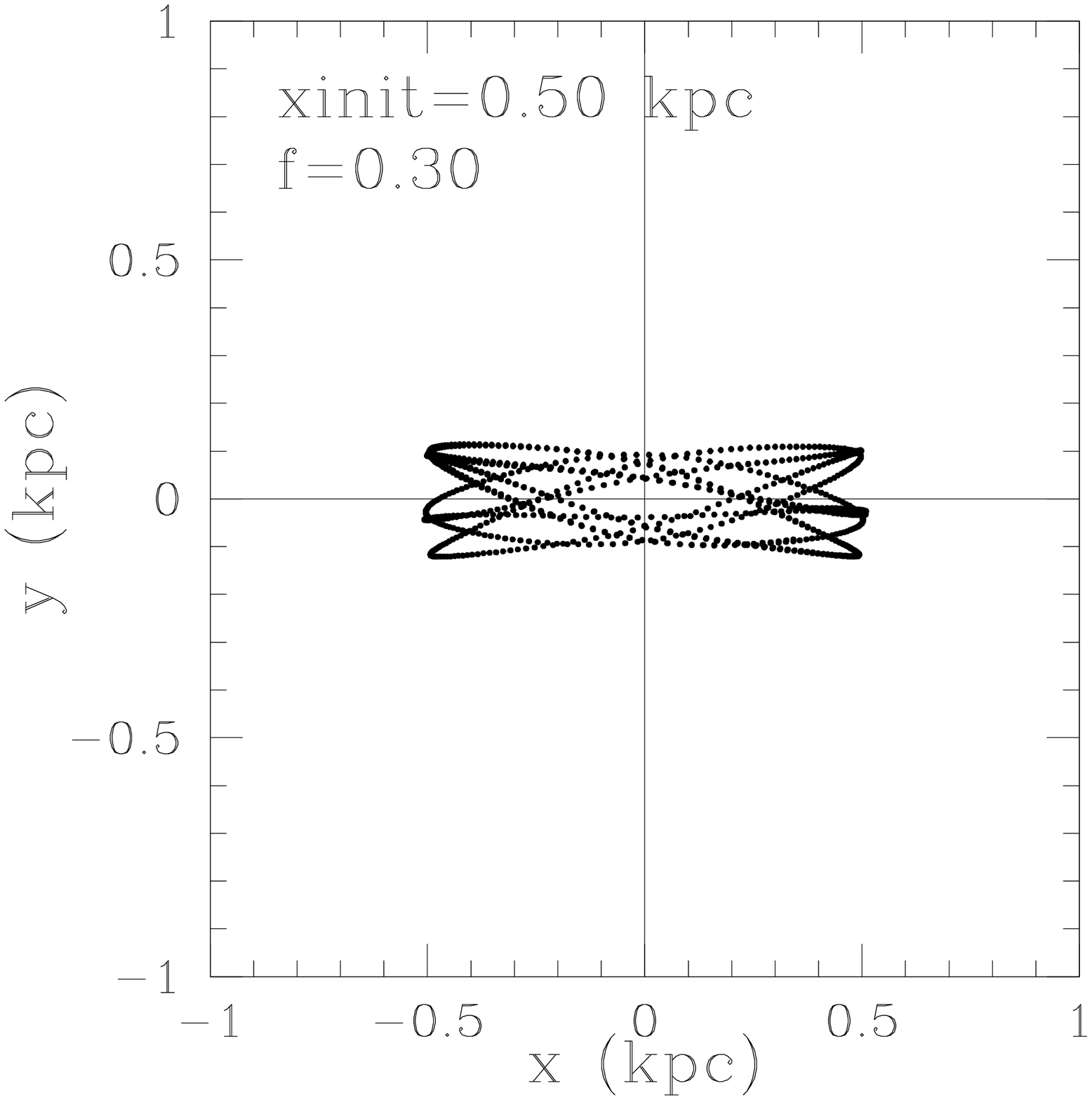,width=4cm}
   \psfig{figure=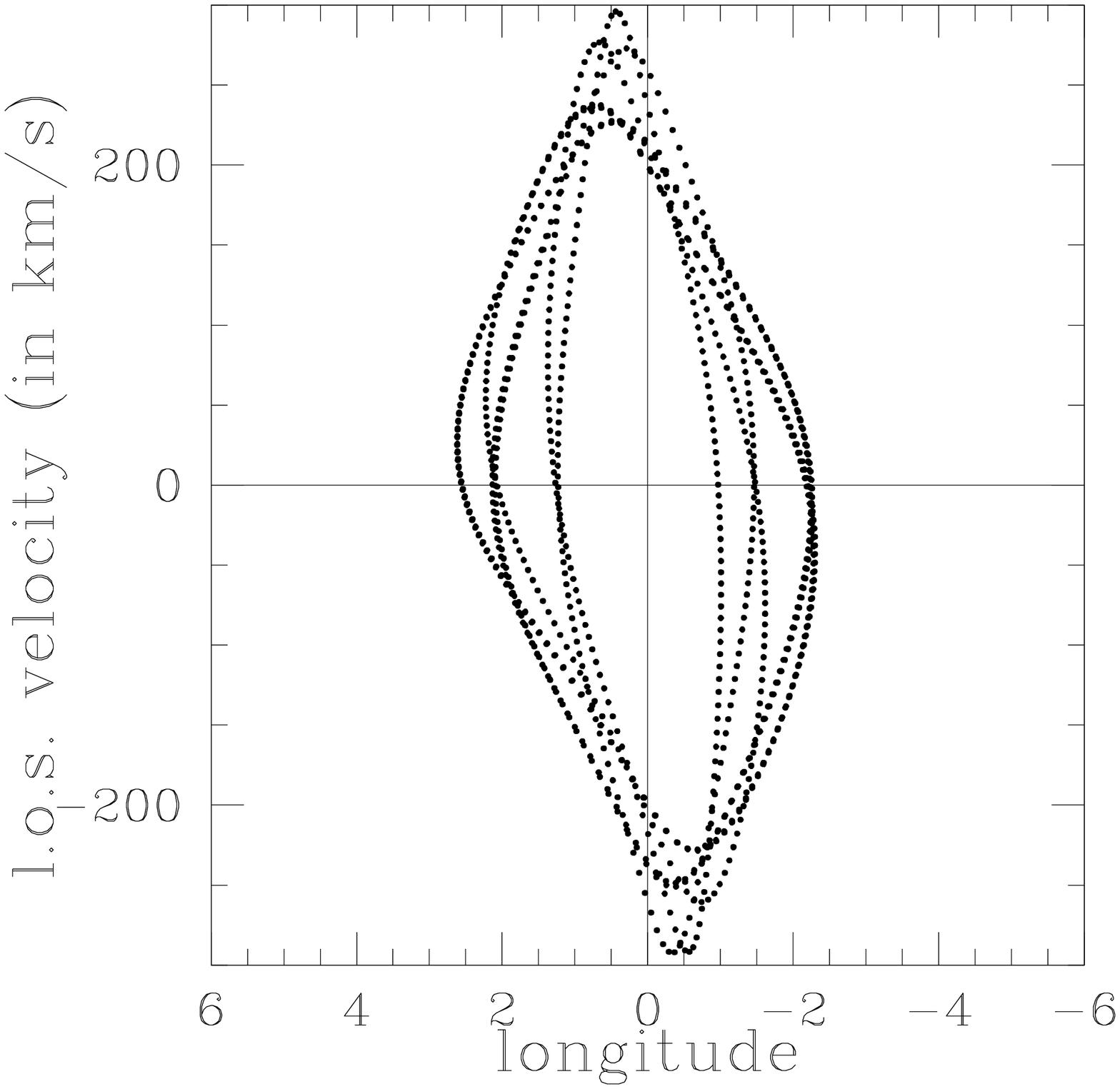,width=4cm}
   \psfig{figure=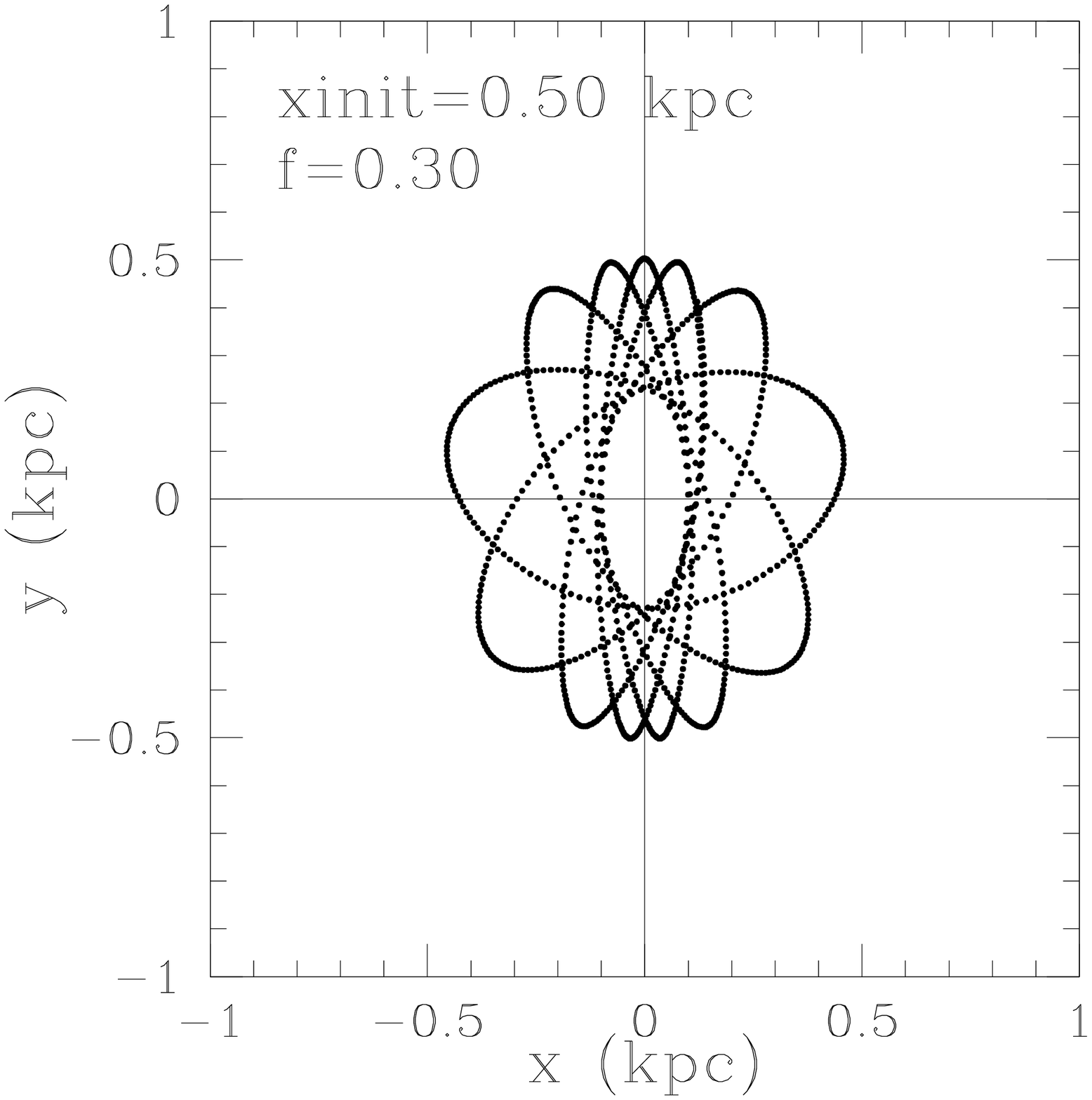,width=4cm}
   \psfig{figure=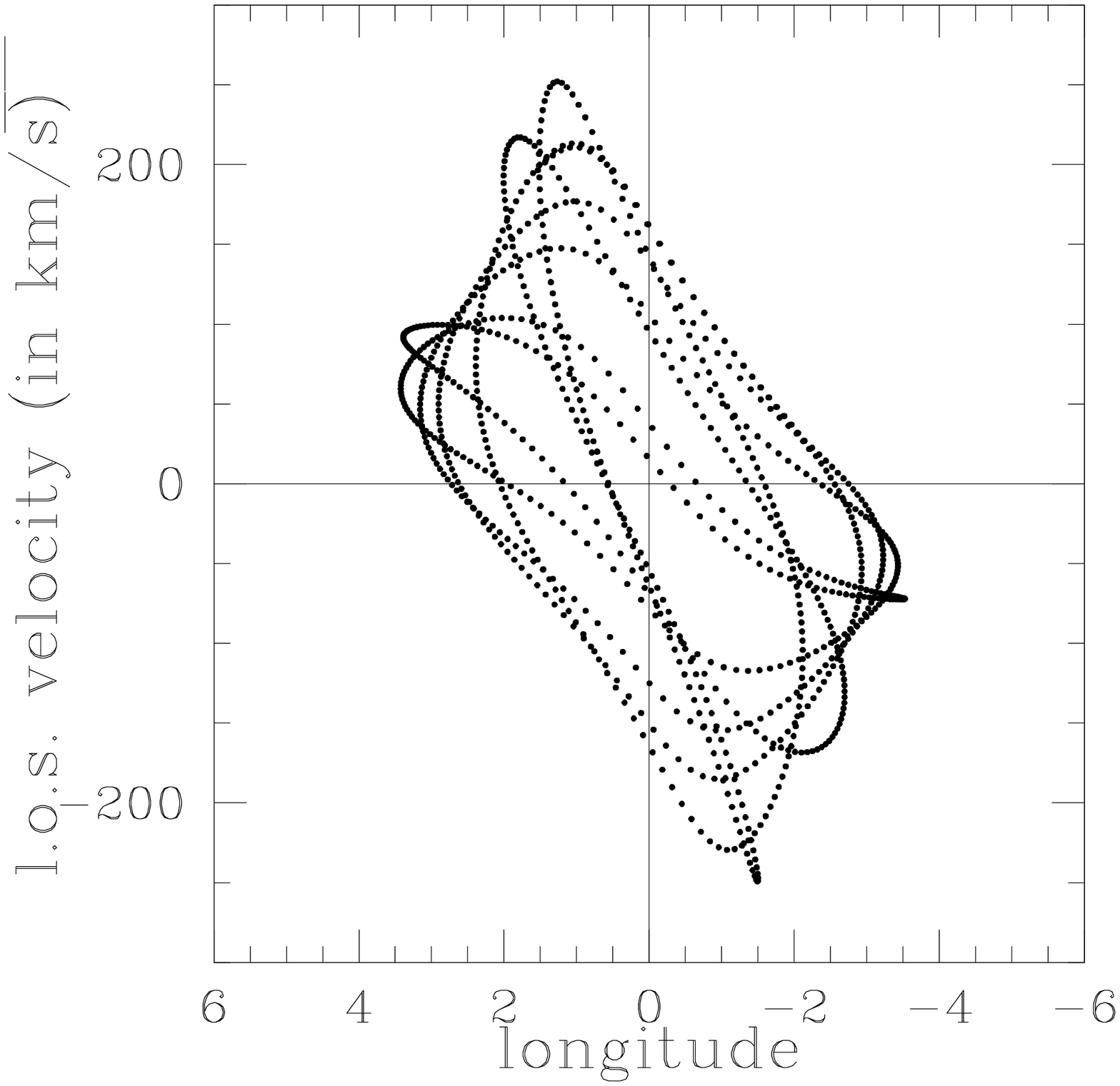,width=4cm}}
   \centerline{\psfig{figure=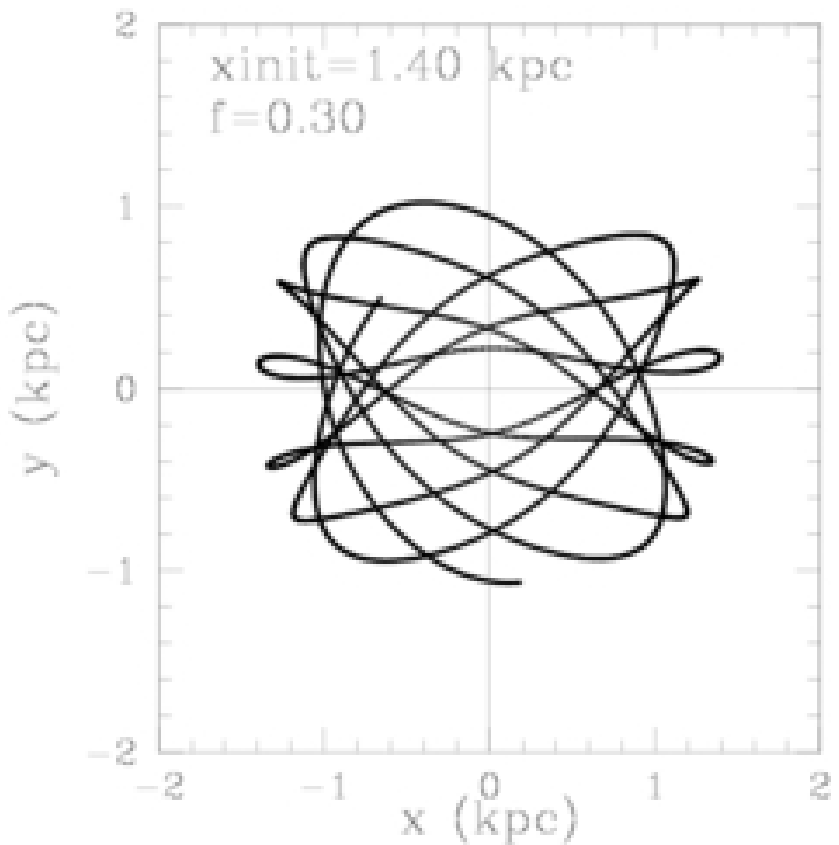,width=4cm}
   \psfig{figure=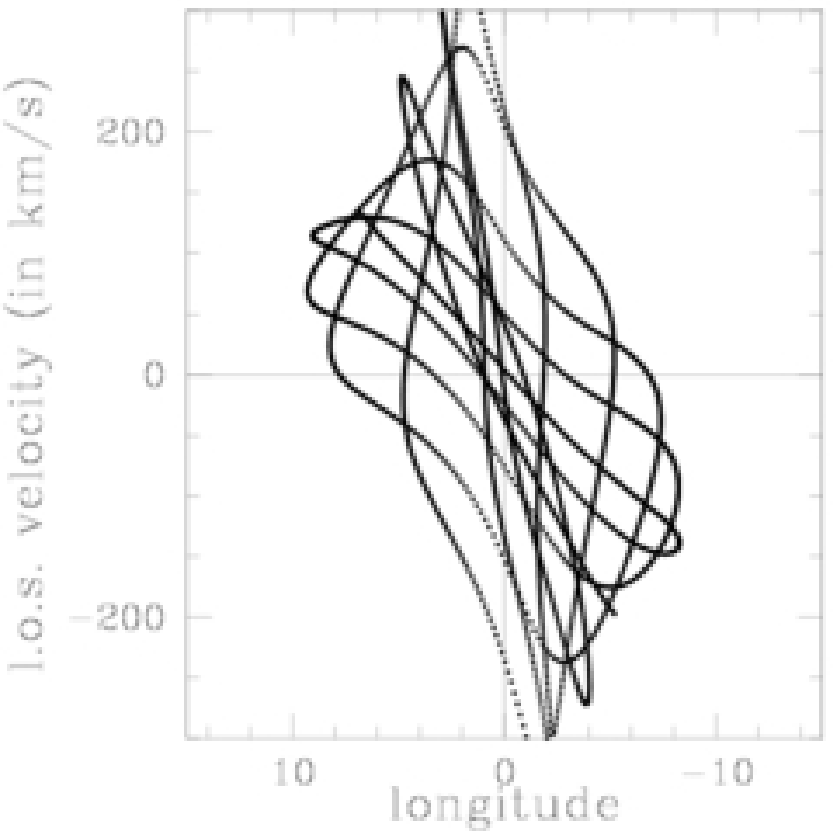,width=4cm}
   \psfig{figure=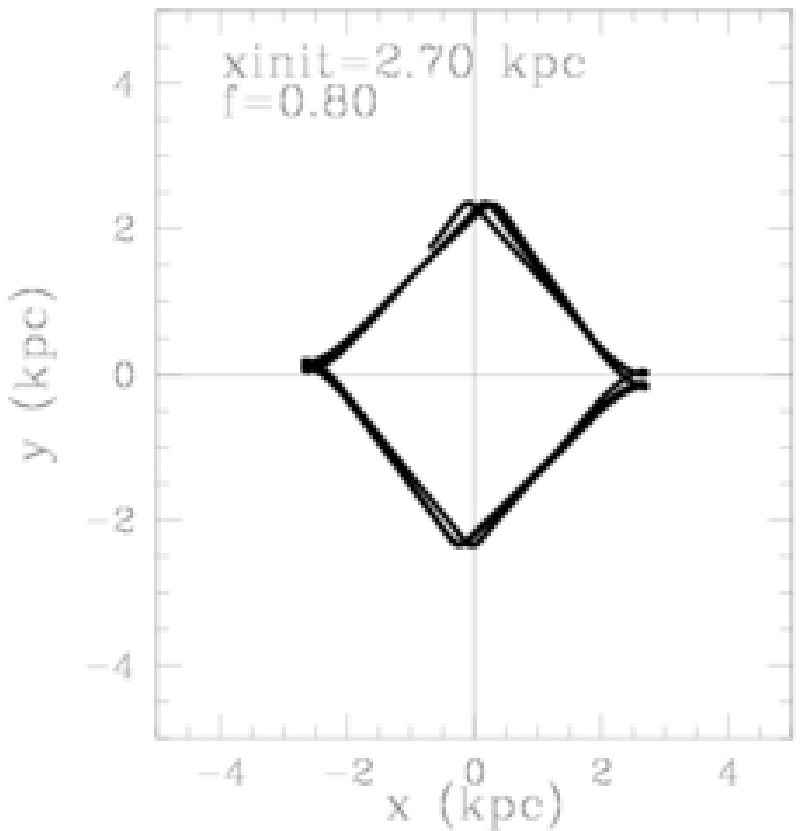,width=4cm}
   \psfig{figure=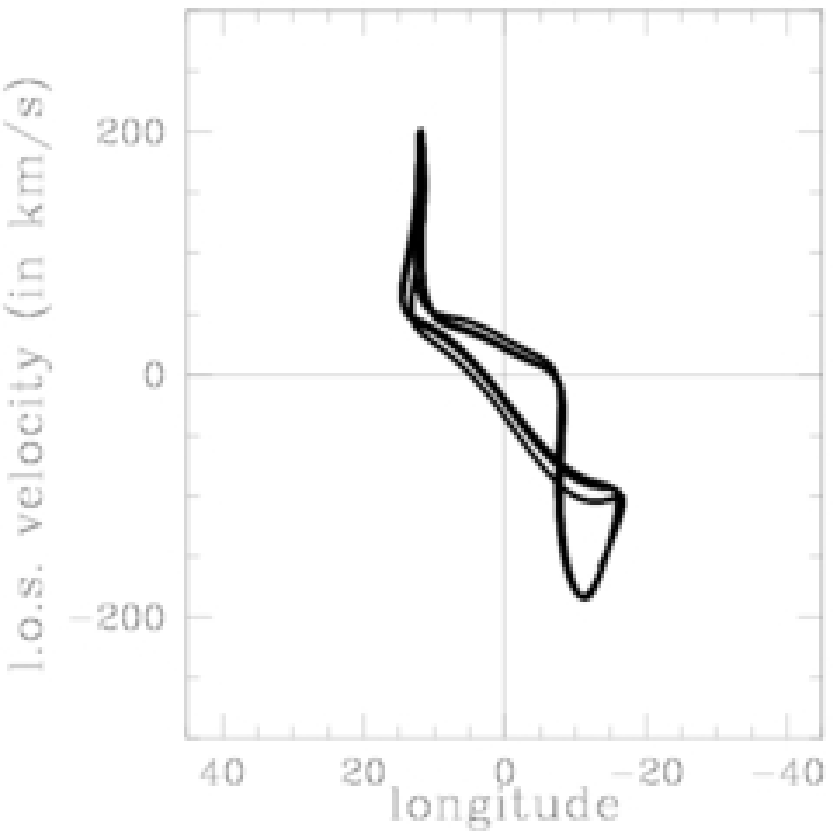,width=4cm}}
   \centerline{\psfig{figure=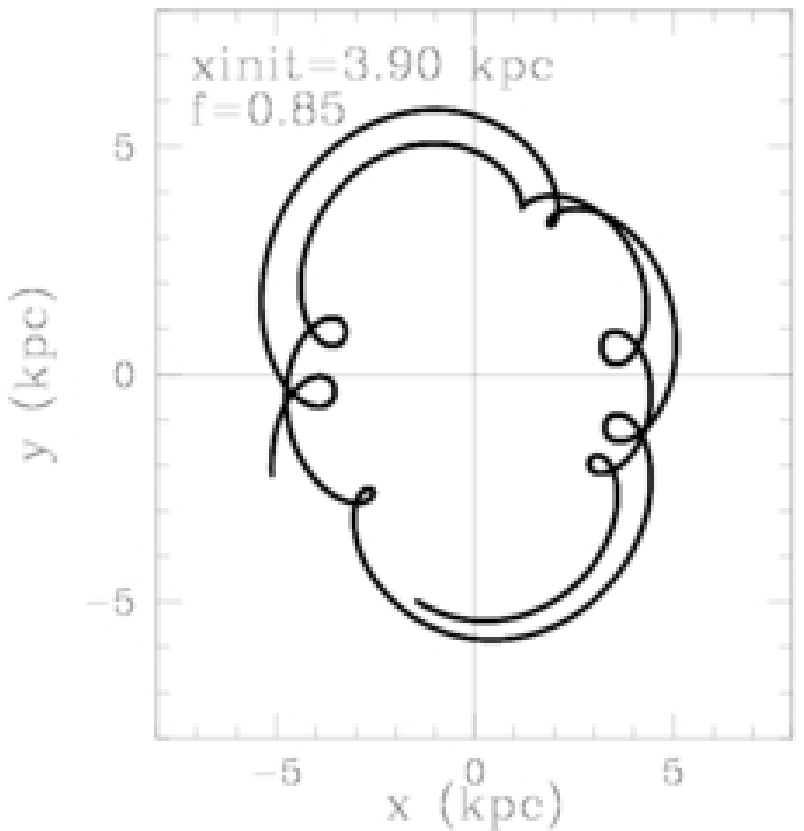,width=4cm}
   \psfig{figure=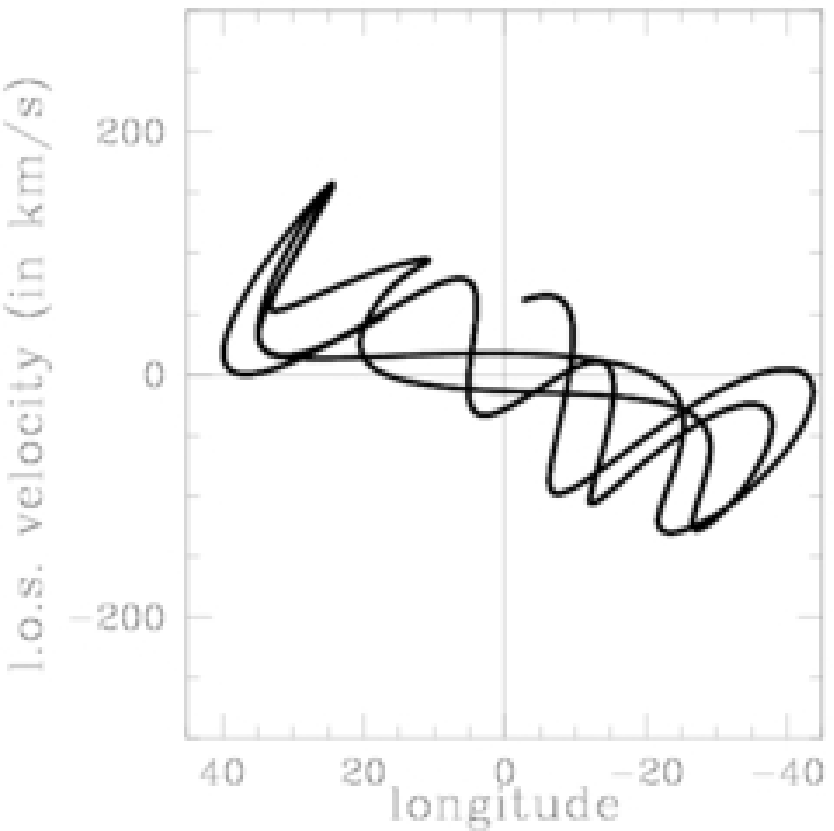,width=4cm}
   \psfig{figure=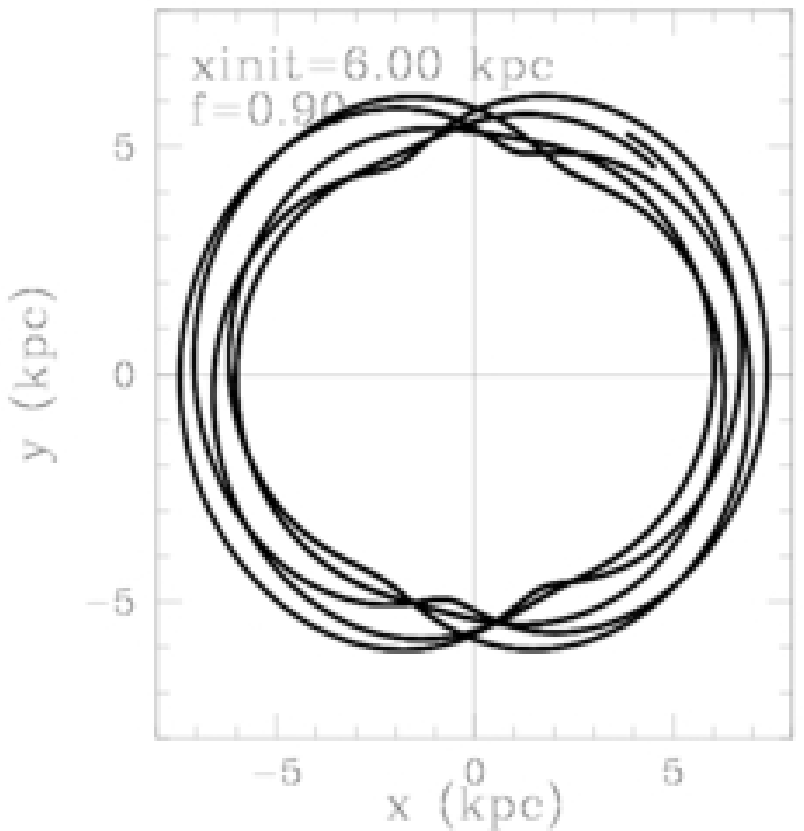,width=4cm}
   \psfig{figure=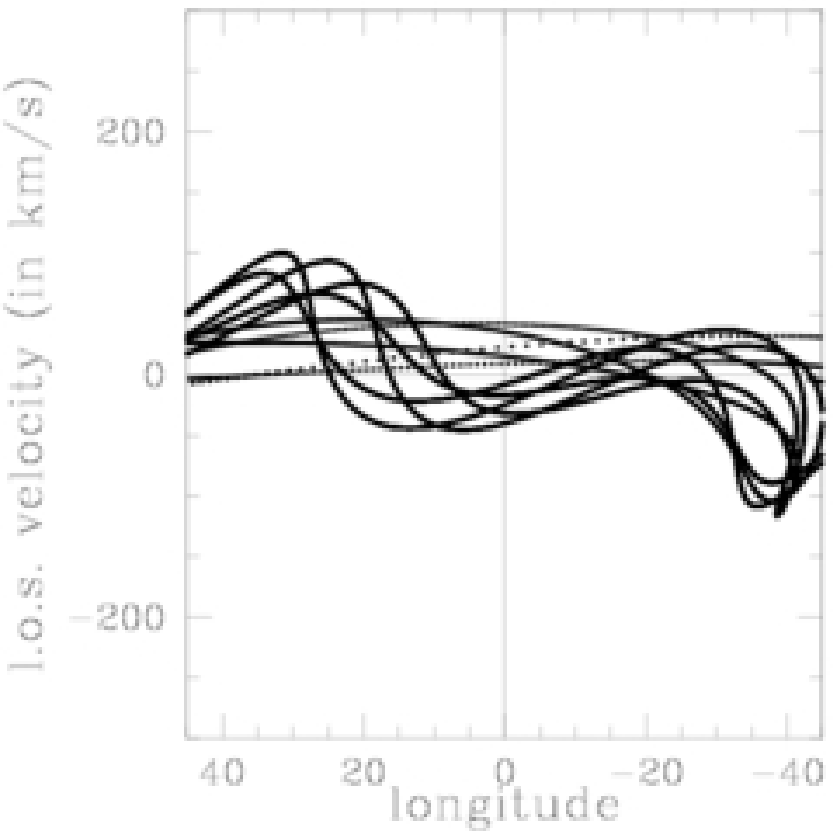,width=4cm}}
   \caption{Six pairs of diagrams, each pair referring to a specific orbit.
     Each diagram contains $2000*$\xinit\ stars (\xinit in kpc) that
     started their journey in the bar at a distance \xinit\ from the
     center under identical initial conditions but at successive times
     separated by 2.5 Mjr. In each diagram the numbers in the upper
     left corner specify the orbit (see text). Of each pair of
     diagrams the one on the left is the distribution in the galactic
     plane and the diagram at the right-hand side is its projection
     onto the \lVd . The horizontal scale varies from one \lVd\ to the
     next.}
   \end{figure*}

\section{Results and discussion}

After some experimentation we decided to use 44 orbits arranged in six
groups. Each orbit is defined by an initial position and an initial
velocity- see Table~3 that contains information on all orbits
(``orb.'') used, on the group (``gr'') to which they were associated
and on the weights (``$w_i$'') that we finally assigned.\\ 

Fig.~6 contains six pairs of diagrams representing the six groups of
stars; for clarity they contain only about 10\% of the points that
have been calculated. The left diagram of each pair shows the orbit
projected on the galactic plane. The diagram on the right gives the
distribution in the \lVd .

\subsection{Choice of weights that produce the best fit to the observations}

We start by filling area 7 that contains 11 OH/IR stars with high
velocities ($\mid $\vlos$\mid > 150$ \kms ) in the forbidden quadrants
Q2 and Q4. It is easily seen in Table~1 that stars of group 1 are
especially needed for this purpose; these are the stars on bar orbits
as discussed above. These stars will also begin to fill areas 3, 4 and
6.  Next consider areas 1 and 2: these will be filled predominantly by
stars on the orbits of groups 5 and 6. The remaining areas (3, 4, 5,
6) have now been partially filled by orbits of groups 1, 5 and 6 and
we use the other groups to complete the population of the remaining
areas. This led us to a final assignment of the weights $w_i$ as given
in Table~3. This table contains a specification of the solution with
the smallest value of $\Sigma\,\chi_i^2$. The first column contains
the number of the orbit and in columns (2) through (5) the orbit is
specified.  Columns (6) through (13) contain the number of maser stars
in each of the seven counting areas and their sum. Column (14) gives
the number of stars on this orbit but outside of the range, i.e stars
for which $\mid l\mid > 45^\circ$.

\begin{table*}
\begin{center}
\begin{tabular}{rcccccrrrrrrrr}\hline
\multicolumn{14}{c}{Table~3}\\
\hline \multicolumn{14}{c}{all orbits, their parameters and
their contribution to the 7 areas}\\
orb&\xinit&\yinit&\vxinit&\vyinit&\multicolumn{7}{c}{predicted number in
each area}&sum&outside\\
&kpc&kpc&\kms&\kms&[1]&[2]&[3]&[4]&[5]&[6]&[7]&&\\
\hline
(1)&(2)&(3)&(4)&(5)&(6)&(7)&(8)&(9)&(10)&(11)&(12)&913)&(14)\\
\hline
&&&&&&&&&&&&&\\
\multicolumn{14}{l}{group 1, $f_1=0.30$ , $w_1= 2.00$}\\
   1&   0.30&   0.00&   0& -54& 0& 0& 3&11& 0& 7& 2& 23&  0\\
   2&   0.40&   0.00&   0& -56& 0& 0& 4& 9& 0& 7& 2& 22&  0\\
   3&   0.50&   0.00&   0& -57& 0& 0& 4& 9& 0& 7& 2& 22&  0\\
   4&   0.60&   0.00&   0& -58& 0& 0& 4& 9& 0& 7& 2& 22&  0\\
   5&   0.70&   0.00&   0& -59& 0& 0& 5& 8& 0& 7& 2& 22&  0\\
   6&   0.80&   0.00&   0& -59& 0& 0& 5& 8& 0& 7& 2& 22&  0\\
\hline
&&&&&&&&&&&&&\\
\multicolumn{14}{l}{group 2, $f_2=0.30 $ , $w_2= 0.10$}\\
   7&   0.00&   0.30&  54&  0&  0& 0& 0& 1& 0& 0& 0&  1&  0\\
   8&   0.00&   0.50&  57&  0&  0& 0& 0& 1& 0& 0& 0&  1&  0\\
   9&   0.00&   0.70&  59&  0&  0& 0& 0& 1& 0& 0& 0&  1&  0\\
  10&   0.00&   0.80&  59&  0&  0& 0& 0& 1& 0& 0& 0&  1&  0\\
  11&   0.00&   0.90&  59&  0&  0& 0& 0& 1& 0& 0& 0&  1&  0\\
\hline
&&&&&&&&&&&&&\\
\multicolumn{14}{l}{group 3, $f_3=0.30$ , $w_3= 0.20$}\\
  12&   1.00&   0.00&   0& -59& 0& 0& 0& 1& 0& 1& 0&  2&  0\\
  13&   1.10&   0.00&   0& -59& 0& 0& 1& 1& 0& 1& 0&  3&  0\\
  14&   1.20&   0.00&   0& -60& 0& 0& 1& 1& 0& 1& 0&  3&  0\\
  15&   1.30&   0.00&   0& -60& 0& 0& 1& 1& 0& 1& 0&  3&  0\\
  16&   1.40&   0.00&   0& -60& 0& 1& 0& 1& 0& 0& 0&  2&  0\\
  17&   1.50&   0.00&   0& -60& 0& 1& 0& 1& 0& 0& 0&  2&  0\\
  18&   1.70&   0.00&   0& -60& 0& 1& 0& 1& 0& 0& 0&  2&  0\\
  19&   1.90&   0.00&   0& -60& 0& 0& 0& 0& 0& 0& 0&  0&  0\\
\hline
&&&&&&&&&&&&&\\
\multicolumn{14}{l}{group 4, $f_4=0.80$ , $w_4= 0.95$}\\
  20&   2.10&   0.00&   0&-160& 0& 7& 0& 3& 0& 1& 0& 11&  0\\
  21&   2.30&   0.00&   0&-160& 0& 7& 0& 3& 0& 1& 0& 11&  0\\
  22&   2.50&   0.00&   0&-160& 0& 7& 0& 2& 0& 1& 0& 10&  0\\
  23&   2.70&   0.00&   0&-160& 2& 7& 0& 1& 0& 1& 0& 11&  0\\
  24&   2.90&   0.00&   0&-160& 1& 6& 0& 2& 0& 1& 0& 10&  0\\
  25&   3.10&   0.00&   0&-160& 2& 6& 0& 2& 0& 1& 0& 11&  0\\
  26&   3.30&   0.00&   0&-160& 3& 5& 0& 2& 0& 1& 0& 11&  0\\
\hline
&&&&&&&&&&&&&\\
\multicolumn{14}{l}{group 5, $f_5=0.85$ , $w_5= 2.00$}\\
  27&   3.50&   0.00&   0&-170& 6&11& 0& 4& 0& 1& 0& 22&  0\\
  28&   3.70&   0.00&   0&-170& 7&10& 0& 3& 0& 2& 0& 22&  0\\
  29&   3.90&   0.00&   0&-170&12& 4& 0& 2& 3& 1& 0& 22&  2\\
  30&   4.10&   0.00&   0&-170&13& 5& 0& 2& 2& 1& 0& 23&  1\\
  31&   4.30&   0.00&   0&-170&15& 4& 0& 2& 0& 1& 0& 22&  0\\
  32&   4.50&   0.00&   0&-170&15& 3& 0& 3& 0& 1& 0& 22&  0\\
\hline
&&&&&&&&&&&&&\\
\multicolumn{14}{l}{group 6, $f_6=0.85$ , $w_6= 3.75$}\\
  33&   4.75&   0.00&   0&-170&24& 5& 0& 4& 5& 2& 0& 40&  3\\
  34&   5.00&   0.00&   0&-170&26& 5& 0& 4& 3& 3& 0& 41&  3\\
  35&   5.25&   0.00&   0&-170&16& 4& 0& 4& 7& 3& 0& 34& 20\\
  36&   5.50&   0.00&   0&-170&18& 2& 0& 3& 7& 1& 0& 31& 24\\
  37&   5.75&   0.00&   0&-170&20& 2& 0& 1& 8& 2& 0& 33& 20\\
  38&   6.00&   0.00&   0&-170&21& 1& 0& 1& 9& 2& 0& 34& 18\\
  39&   6.25&   0.00&   0&-170&23& 0& 0& 1& 9& 1& 0& 34& 16\\
  40&   6.50&   0.00&   0&-170&24& 0& 0& 1& 9& 1& 0& 35& 15\\
  41&   6.75&   0.00&   0&-170&24& 0& 0& 1& 8& 1& 0& 34& 16\\
  42&   7.00&   0.00&   0&-170&24& 0& 0& 2& 6& 2& 0& 34& 21\\
  43&   7.25&   0.00&   0&-170&21& 0& 0& 2& 6& 2& 0& 31& 27\\
  44&   7.50&   0.00&   0&-170&19& 0& 0& 2& 6& 2& 0& 29& 33\\
\hline
\end{tabular} \label{tab:allebanen}
\end{center}
\end{table*}

   \begin{figure*}[t!]\label{fig:grotediagram}
   \centering
   \centerline{\psfig{figure=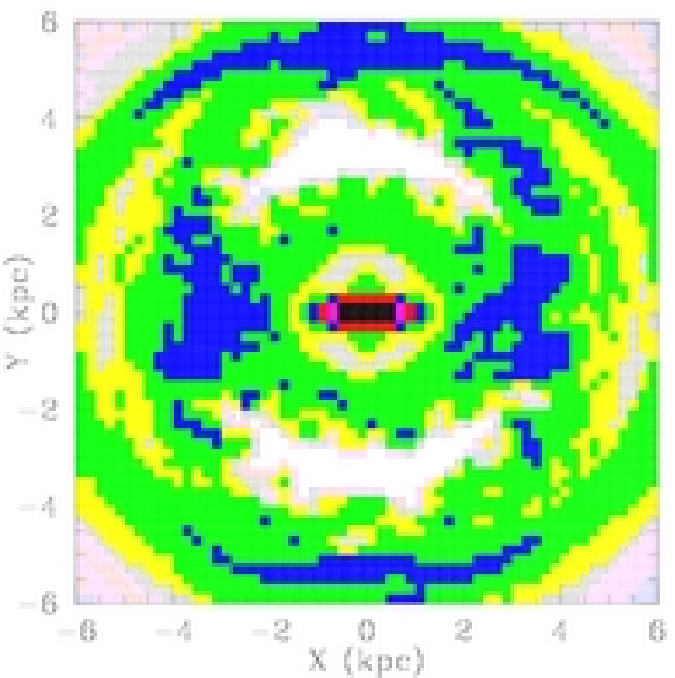,width=6cm}
   \psfig{figure=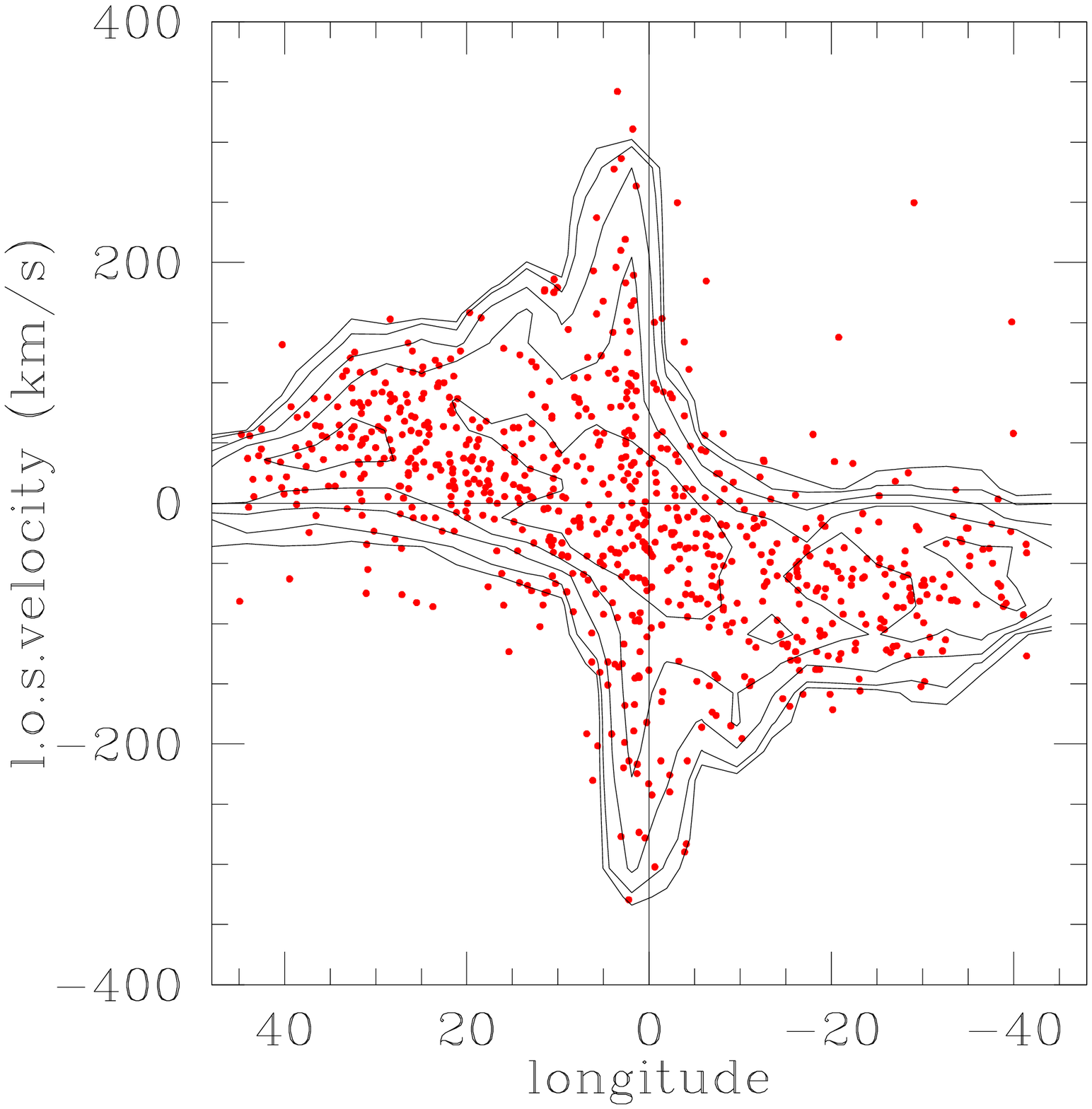,width=6cm}}
    \caption{On the left our ultimate contour diagram showing the
      reconstructed distribution of the maser stars in an area of
      $12\times 12$ kpc$^2$ centered on the GC. Notice the bar like
      distribution near the center and the voids at about $(x,y)=(
      0.0$ kpc,$\, \pm 3$ kpc). The average stellar density can be
      read off from the colors: black: 110 maser stars per kpc$^2$,
      magenta: 56; red: 24, blue 8.9, green 5.4, yellow 2.6; other
      colors indicate less than 1.0 star per kpc$^2$; the white area
      does not contain a single star. On the right the corresponding
      diagram of longitude versus \los -velocity of the OH/IR stars.
      The contours in black illustrate our prediction and the red dots
      represent the observed distribution of the OH/IR stars.}
    \end{figure*}
    
    Using the orbits and weights specified in Table~3 we find the
    total number of stars in each of the seven areas; see Table~2. The
    first row in Table~2 gives the total number of stars predicted,
    the second row gives the number as observed. In rows three and
    four we show for each area the two $\chi^2$-values that follow from
    the comparison of predictions and observations (see above).\\ 

\begin{table}
\begin{center}
\begin{tabular}{r|ccccccc|c}\hline
\multicolumn{9}{c}{Table~2}\\
\multicolumn{9}{c}{number of stars in each of the seven areas}\\
\hline area: &  1 & 2& 3& 4& 5& 6& 7& sum\\
\hline
pred.& 336 & 104 &  28 & 123 & 88 & 82 & 12 & 773 \\
obs.& 355 & 108 &  19 & 104 & 92 & 82 & 11 & 771 \\
\hline
$\chi^2_1$ & 1.1 & 0.2 & 2.9 & 2.9 & 0.2 & 0.0 & 0.1& 7.3\\
$\chi^2_2$ & 1.0 & 0.1 & 4.3 & 3.5 & 0.2 & 0.0 & 0.1& 9.2\\
\hline
\end{tabular}\label{tab:somingebieden}
\end{center}
\end{table}

\subsection{The distribution of maser stars in the Inner MWG}

An important result of our simulation are the best weights, $w_i$,
that we assigned to each group of orbits. This set of weights and the
properties of each orbit represent together the distribution function
as known from statistical mechanics. We may use this function to
calculate the spatial distribution of the stars. The left diagram in
Fig.~7 shows the result obtained after smoothing the somewhat noisy
image. The map shows two striking features. First there is the
rectangular shaped distribution of stars near the GC: the galactic
bar.  The existence of the bar is required by the stars in area 7
(high velocities near the GC and in the forbidden quadrants). The 30
OH/IR stars in areas 3 and 7 belong to the bar; they travel up and
down the bar. The second remarkable feature is the presence of two
empty regions, the voids, one with the shape of the crescent of the
Moon at first quarter, the other with that of the Moon at last
quarter. The existence of these voids is described in Sect.~5.3.2;
there it is also argued that the observations show the absence of
stars in these crescents; the crescents are voids in the maser star
distribution.  The relative size of the voids is determined by the
strength of the bar, i.e. by the value of the parameter $\epsilon$ in
Eq.  (\ref{eq:Phipotbar}). The voids disappear when
$\epsilon\rightarrow 0$.\\ 

We point out that our value for co-rotation, 3.3 kpc agrees with the
radius of the molecular ring. It is likely that there is a relation
between the presence of the bar inside 3.3 kpc and the (almost)
complete absence of ISM inside of this distance- with the exception of
the nuclear disk in the immediate neighbourhood of SgrA; below we
suggest a mechanism that may cause a rapid transport of ISM from the
molecular ring to the very center of the MWG.\\ 

The structure shown in Fig.~7 is similar to the figure derived by one
of us in an earlier paper (Sevenster, 1999b).\nocite{seven:99b} She
derived her conclusions on very different arguments, namely on a
morphological study of the distribution of the stars in a diagram of
longitude versus latitude.\\ 

\subsection{Is our representation of the \lVd s by orbits unique?}

We have obtained a set of orbits that fit the observations quite well
and this led us to the reconstruction of the distribution of maser
stars shown in Fig.~7. There remains at least one important question:
can other sets be found that predict \lVd s equally well and yet give
a very different distribution of the maser stars in the galactic
plane?\\ 

We started from a large library of orbits from which we selected a
limited group of orbits that might be significant. We concluded
earlier that outside of the radius of the molecular ring, 3 to 3.5
kpc, the orbits have high angular-momentum as is appropriate for the
galactic disk. Inside 3 kpc the velocity distribution changes, as the
orbits become more and more elongated.  These considerations are
reflected in the list of orbits shown in Table~3. Once this selection
of orbits had been made the assignment of the weights followed more or
less straightforwardly. We thus do not expect that a qualitatively
very different solution can be found with the same low value of
$\Sigma \chi_i^2$.\\ 

Consider now the values of the two most uncertain free parameters,
$\epsilon$, the strength of the bar-like part of our gravitational
potential, and $\phi_\odot$, the angle between the Sun-GC line and the
bar. We adopted $\epsilon=0.11$, partially on intuition and partially
because it is in the same range as the potential used by Bissantz et
al. in their analysis of the ISM distribution in the inner MWG. We
redid our analysis assuming $\epsilon=0.05$ and found a set of weights
by which $\Sigma \chi_i^2= 10$ but the overall shape of the
distribution looked suspect. We concluded that a value of
$\epsilon=0.05$ is barely acceptable and that a lower value than
$0.05$ is ruled out. Taking $\epsilon=0.2$ led to predicted \lVd s
that contained too many sources in area 7. Our estimate is thus that
$\epsilon$ has a value between 0.05 and 0.2, but the uncertainty in
this value we cannot specify quantitatively. With respect to
$\phi_\odot$ we have experimented with values of $20^\circ$ and
$40^\circ$ and find easily equally satisfying solutions. From star
count data in the mid-infrared taken with the SPITZER satellite
Benjamin et al. (2005) find a value of $\phi_\odot= 45^\circ$.\\ 

In this paper our approach has been to chose a galactic potential that
contains a rotating bar, followed by orbit calculations and then by a
selection of those orbits that predict as well as possible the
observed \lVd s. Given this approach we think that the end result is
the best possible, and actually that it is quite good. There are
however enough uncertainties in our approach (how reliable is the
potential; are there many stars below the detection limits of the
maser surveys and are the observed \lVd s sufficiently representative
of the true distribution?) to leave some lingering
doubts- even among the authors of this paper.\\

\subsection{Speculation about the ``3 kpc arm''}

   \begin{figure*}[t!]\label{fig:sim3kpc}
   \centering
   \includegraphics[width=8cm]{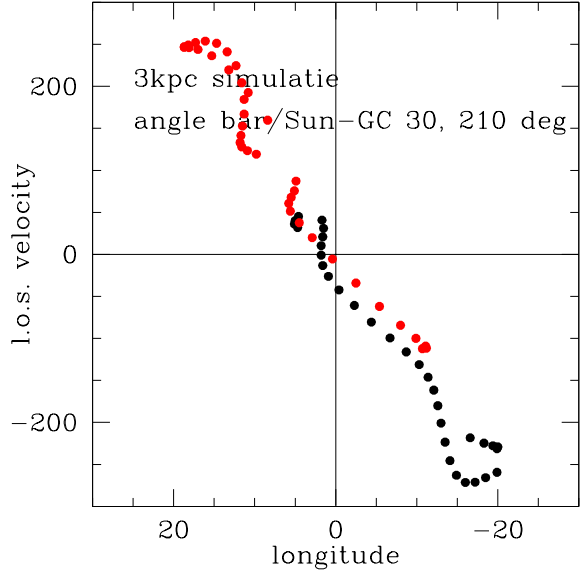}
   \caption{The relation between longitude and \vlos\ for projectiles
     launched from the points of contact B1 (black symbols) and B2
     (red symbols). We observe $l$ and \vlos\ of each projectile when
     the bar makes an angle of $30^\circ$ or of $210^\circ$ with the Sun-GC
     line. The time between two launches is approximately 5 Myr.}
   \end{figure*}

In the quite successful interpretation of the galactic survey of
the 21cm line in the 1950's there remained one major puzzle, an
easily recognized feature that has never been properly explained.
This is the so-called "3kpc-arm", an object that is almost
certainly not a spiral arm (i.e. it is not a ripple on the
galactic distribution of the ISM) and may not be at 3 kpc either.
At $l=0^\circ$ it is seen as an absorption line against Sgr A with
a \los -velocity of -50 \kms ; hence it must lie between the Sun
and the GC. This high \los\ velocity has never been explained
satisfactorily, although many attempts have been made. There may
be a counterpart to the 3kpc-arm at the other side of the GC; it
is seen as an elongated feature in HI and CO \lVd s with a
\los-velocity of +135 \kms .\\ 

Sevenster (1999)\nocite{seven:99b} suggested that the arm has its
origin near one of the two points where the bar meets its corotation
radius. Matter might "leak" through this and then fall inwards. In
this model the 3kpc arm is a flow of matter from this corotation point
inward that feeds gas to the clouds close to the GC, e.g. such clouds
as Sgr B2.\\

We calculated orbits for stars. Clouds of ISM will follow the same
orbits until they collide with other clouds. Extended structures of
the ISM will face another problem: during a ballistic orbit some parts
of the structure will be forced by the gravitational force to contract
faster than the speed of sound and shocks will appear that may convert
the kinetic energy of large-scale motions into thermal energy, i.e.
into heat. We have checked this and find that in most of our ballistic
orbits such rapid internal motions of 5 to 10 \kms\ occur during short
periods. We suggest that the shock waves in the ISM and the
consequential lowering of the kinetic energy of the gas may give the
explanation for the low density of the ISM inside CR.\\ 

This conclusion led us to the following simulation. The points where
the bar crosses the co-rotation circle will be named B1 and B2, with
B1 being (at present) closer to the Sun than B2. At successive times,
separated by a fixed interval $\Delta\, t$, a projectile, i.e. a cloud
of ISM or a star, is ejected from B1 or B2 into the inner MWG with an
initial velocity $\vec{\dot r_0}$. We follow all projectiles along
their ballistic orbits in the same way as we followed the stars. When
the clouds/projectile have completed half a revolution around the GC
they dissolve. When the Sun and the bar reach their present mutual
position we calculate the longitude and \vlos\ of all objects.\\ 

Fig.~8 shows the velocities and longitudes of the ``projectiles'' shot
from contact point B1 (black) or B2 (red) and observed from the Sun
when the angle between the Sun-GC line and the central line of the bar
equals $30^\circ$ or of $210^\circ$. This is a crude result: more
simulations are needed but we think that our approach is promising and
that the 3kpc arm may well be a channel through which gas is
transported from the molecular ring to the galactic center area. A gas
dynamical study is required to further develop this suggested
explanation of the 3kpc-arm.\\ 

\subsection{Conclusions and a few concluding remarks}

The following conclusions from our study we find the most
important:
\begin{enumerate}
\item A qualitative comparison of the \lVd s of the ISM and of the
  maser stars shows overall agreement: stars and gas have the same
  overall distributions in space and in velocity. Small but
  significant differences can be attributed to the fact that the gas
  is on dissipative orbits and the maser stars are not.
\item The shape of the velocity distribution of the maser stars
  changes when one moves inward along the galactic plane: outside 3.3
  kpc the distribution is peaked around the circular velocity; inside
  this, i.e. inside of the molecular ring, radial motions become
  increasingly important.
\item Qualitative comparison between the \lVd s of the ISM and of the
  maser stars leads to the demarcation of seven areas in the \lVd s.
\item Adopting a potential that is predominantly axially symmetric but
  contains a weak, rotating bar we find a set of orbits that explains
  the distribution of longitudes and \los -velocities.
\item Inside 2 kpc from the GC the stars are on bar orbits.
\item We also find two crescent-like voids where no star is found.
  These voids are caused by the co-rotation resonance.
\item The distribution of the maser stars as derived in this paper
  compares remarkably well to that presented by Sevenster (1999),
  based on partially the same data but on a very different type of
  argument.
\end{enumerate}

Some concluding remarks \begin{enumerate}
\item The analysis in this paper is 2D and it is based on a simple
  galactic potential. One of us (GvdV) has started a larger 3D
  analysis.
\item It is technically possible to measure the proper motions of the
  SiO maser stars by interferometry. This will be a great step
  forward..
\item The 11 OH/IR stars in area 7 of the \lVd\ are moving up and down
  the bar. Their age is between 1 and 5 Gyr and thus they set a limit
  on the time of formation of the bar.
\end{enumerate}

\begin{acknowledgements} M.M.'s research has been made possible by a 
  Grant from NOVA, the Netherlands Research School in Astronomy. We
  thank the referee, Anders Winnberg, for his fair and critical
  comments.

\end{acknowledgements}

\bibliographystyle{aa}
\bibliography{4480bib}

\end{document}